\RequirePackage{fix-cm}
\documentclass[smallextended]{svjour3}  
\smartqed  
\usepackage{graphicx}
\usepackage{listings}
\usepackage{hyperref}
\usepackage{amssymb}
\usepackage{tablefootnote}
\usepackage{amsmath,amsfonts}
\usepackage{graphicx}
\usepackage{breakurl}
\usepackage{textcomp}
\usepackage[utf8x]{inputenc} 
\usepackage{xcolor}
\usepackage{blindtext}
\usepackage{amsmath}
\usepackage{algorithm}
\usepackage[misc]{ifsym}
\usepackage{algpseudocode}
\usepackage[most]{tcolorbox}
\newcommand*{\affaddr}[1]{#1} 
\newcommand*{\affmark}[1][*]{\textsuperscript{#1}}
\lstdefinestyle{pybox}{
  language=Python,
  basicstyle=\ttfamily\scriptsize,
  keywordstyle=\bfseries\color{blue!70!black},
  stringstyle=\color{orange!80!black},
  commentstyle=\itshape\color{gray!70!black},
  numbers=none,
  xleftmargin=2pt,
  framexleftmargin=2pt,
  breaklines=true,
  showstringspaces=false,
  tabsize=4,
  columns=fullflexible,
  keepspaces=true,
}
\usepackage{pifont}
\usepackage{url}
\usepackage{multirow}
\usepackage[caption=false]{subfig}
\usepackage{ifthen}
\newboolean{showcomments}
\setboolean{showcomments}{true}
\ifthenelse{\boolean{showcomments}}
 { \newcommand{\mynote}[2]{
      \fbox{\bfseries\sffamily\scriptsize#1}
        {\small$\blacktriangleright$\textsf{\emph{#2}}$\blacktriangleleft$}}}
        { \newcommand{\mynote}[2]{}}

\begin{document}

\title{Empirical Study for Structured Output Control in LLMs for Software Engineering}

\titlerunning{Empirical Study for Structured Output Control of LLM for SE}        

\author{Yewei Song\textsuperscript{1}         \and Prateek Rajput\textsuperscript{1} \and
        Tiezhu Sun\textsuperscript{2} \and Saad Ezzini\textsuperscript{3} \and Tegawendé F. Bissyandé\textsuperscript{1} \and Jacques Klein\textsuperscript{1}
}


\institute{\Letter $\;\;$  Yewei Song*\\     \email{yewei.song@uni.lu}     \\\\      
Prateek Rajput \at
               \email{prateek.rajput@uni.lu}
         \\\\    
           Tiezhu Sun \at
              \email{tiezhu.sun@list.lu}
           \\\\
           Saad Ezzini \at
              \email{saad.ezzini@kfupm.edu.sa}
              \\\\
              Tegawendé F. Bissyandé \at
              \email{tegawende.bissyande@uni.lu}
         \\\\
         Jacques Klein \at
               \email{jacques.klein@uni.lu}
         \and
              \affaddr{\affmark[1]University of Luxembourg}\\
              \affaddr{\affmark[2]Luxembourg Institute of Science and Technology}\\
              \affaddr{\affmark[3]King Fahd University of Petroleum and Minerals, Saudi Arabia}\\
              \affaddr{\affmark[*]Corresponding author}
}

\date{Received: date / Accepted: date}

\maketitle

\begin{abstract}
LLM-generated outputs in software engineering rarely exist in isolation. They must plug into toolchains, APIs, and data pipelines that impose strict, often organization-specific structural contracts. A semantically correct output that violates the expected format is, from the consuming system's perspective, indistinguishable from a wrong answer, making structural fidelity an operational prerequisite for deploying LLMs in practice. Yet current models routinely produce syntactically invalid or structurally non-compliant outputs. Unlike encoders, autoregressive decoders generate text token-by-token with a local rather than global focus, amplifying structural fragility whenever the target format deviates from familiar training distributions.

We present a systematic evaluation of structural reliability across four representative SE tasks, categorizing failures into syntax, structural, and semantic errors. We benchmark ways of mitigation targeting the decoder: grammar-constrained decoding, regex-based validation, and a strict template-driven control (Template Token Match Generation, TTMG) to isolate the sources of these failures. TTMG nearly eliminates syntax errors, yet substantial structural and semantic errors persist, demonstrating that the core bottleneck lies beyond syntax formatting. A detailed case study further illustrates how residual errors cascade in downstream workflows. Our findings show that current structure-enforcing tools are necessary but insufficient, and highlight the need for approaches that jointly ensure structural fidelity and semantic correctness in LLM-driven workflows.


\end{abstract}

\section{Introduction}


Large language models (LLMs) have demonstrated strong performance across a wide variety of software engineering (SE) tasks, ranging from code generation and debugging to API synthesis and configuration authoring \cite{fan2023large}. However, many of these tasks require outputs that conform to strict structural, syntactic, or semantic rules—such as valid JSON, SQL queries, or specific function signatures. When such structure is required, the reliability of LLMs often falters. Even advanced models may generate outputs that are syntactically invalid, structurally inconsistent, or incompatible with downstream systems. These issues are not merely cosmetic; a small formatting error, such as a missing bracket or extraneous comma, can render an otherwise correct output unusable, potentially breaking automated pipelines or causing critical failures. 

In practice, LLM-generated outputs are rarely consumed in isolation. They must integrate into existing toolchains, APIs, and data pipelines that impose strict local constraints: a generated API call must conform to the specific schema of the target service, a configuration file must match the format expected by the deployment framework, and a database query must respect the conventions of the local engine. These constraints are inherently \emph{situational}—they vary from organization to organization, from project to project, and even from endpoint to endpoint—shaped by legacy systems, internal conventions, and domain-specific requirements that are unlikely to be well represented in any model's training data \cite{li2024large,uygun2024local,wang2025ai}. An output that captures the correct intent but violates the expected structure is, from the perspective of the consuming system, indistinguishable from a wrong answer.

\begin{figure}
    \centering
    \includegraphics[width=\linewidth]{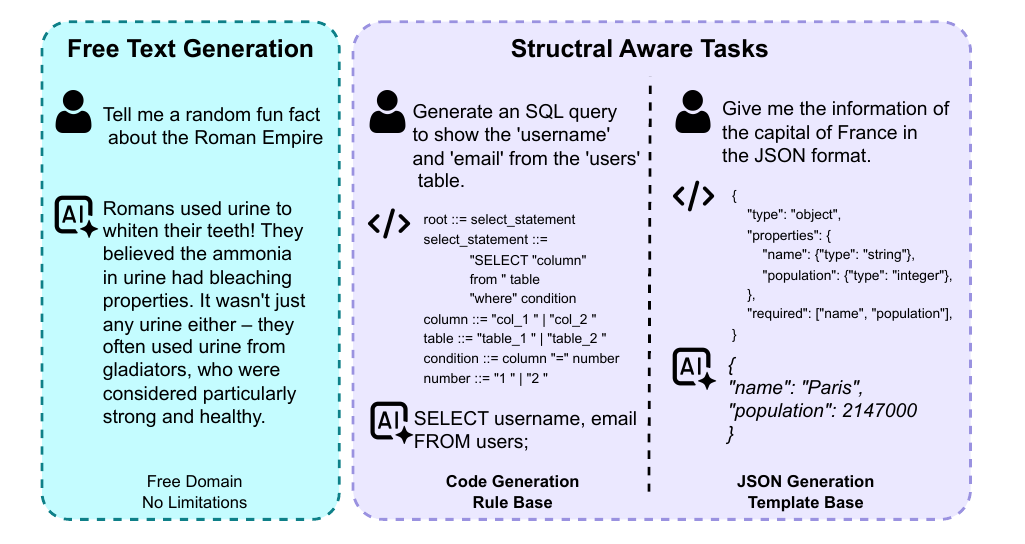}
    \caption{Tasks for Structural Output, differs to free text generation.}
    \label{fig:tasks}
    \vspace{-1em}
\end{figure}

Figure~\ref{fig:tasks} illustrates this contrast by comparing free-text generation to structure-aware tasks. In free-text settings, such as answering open-ended factual questions, LLMs enjoy flexibility and impose no strict formatting constraints. In contrast, structure-aware tasks—like generating SQL queries or JSON objects—require adherence to specific schemas, rules, or templates. For instance, when generating an SQL query, the output must follow a defined grammar to be executable. Similarly, producing a JSON object that satisfies a schema involves including all required fields with the correct types. While the LLM's fluency enables it to produce plausible-looking outputs in both cases, structured tasks demand a higher level of precision and constraint adherence that current models frequently fail to guarantee. Structural adherence is therefore not a secondary concern but a fundamental requirement for practical utility. A structurally malformed output does not merely fail silently. It can cascade through automated pipelines triggering downstream errors, corrupting data, or exposing sensitive information \cite{bazinska2025breaking,yang2025structeval}. This risk is compounded by the fact that LLM-generated outputs are increasingly consumed programmatically rather than reviewed by a human before execution \cite{mitchell2025fully}, leaving little room for manual correction.

Consider a typical scenario: a software engineer asks an AI assistant to generate a JSON configuration snippet for deploying a new microservice. The assistant confidently returns a block of JSON—yet a stray comma or missing closing brace makes it invalid, causing the deployment to fail. This example encapsulates the broader challenge: ensuring LLM outputs are not only semantically relevant but also structurally correct and reliably usable in real-world SE workflows. This work stems from the process of implementing \textbf{AI-driven services} in a large bank, where a small formatting error, such as a missing bracket or extraneous comma in a generated API call, can render an otherwise correct output unusable, potentially breaking automated pipelines or causing severe data leakage. 

{
\begin{figure}[t]
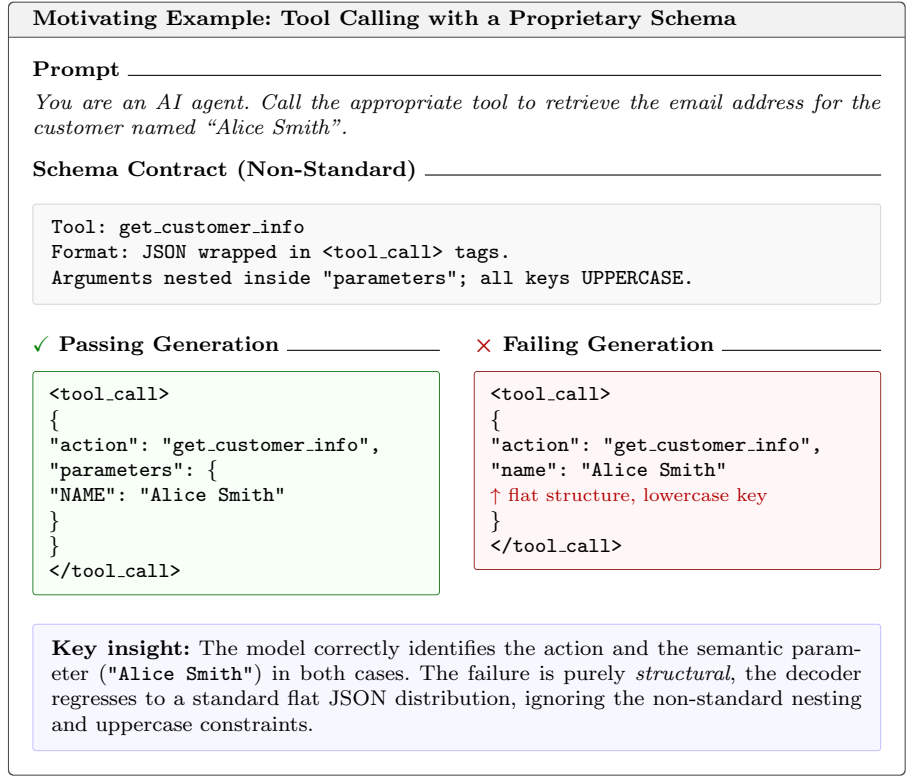

\begin{tcolorbox}[
  colback=white, colframe=black!70, boxrule=0.4pt,
  arc=2pt, outer arc=2pt,
  left=6pt, right=6pt, top=6pt, bottom=6pt,
  fonttitle=\small\bfseries,
  title={Motivating Example: Tool Calling with a Proprietary Schema},
  coltitle=black, colbacktitle=gray!10, boxrule=0.4pt
]

\small

\noindent\textbf{Prompt} \hrulefill
\smallskip

\noindent\textit{You are an AI agent. Call the appropriate tool to retrieve the email address for the customer named ``Alice Smith''.}

\medskip
\noindent\textbf{Schema Contract (Non-Standard)} \hrulefill
\smallskip

\begin{tcolorbox}[
  colback=gray!5, colframe=gray!40, boxrule=0.3pt,
  arc=1pt, left=4pt, right=4pt, top=3pt, bottom=3pt
]
\ttfamily\footnotesize
\noindent Tool: \textbf{get\_customer\_info}\\
Format: JSON wrapped in \texttt{<tool\_call>} tags.\\
Arguments nested inside \texttt{"parameters"};\ all keys \textbf{UPPERCASE}.
\end{tcolorbox}

\medskip

\begin{minipage}[t]{0.48\linewidth}
\noindent\textbf{\textcolor{green!50!black}{\checkmark}\ Passing Generation} \hrulefill

\begin{tcolorbox}[
  colback=green!3, colframe=green!40!black, boxrule=0.3pt,
  arc=1pt, left=3pt, right=3pt, top=3pt, bottom=3pt
]
\ttfamily\footnotesize
<tool\_call>\\
\{\\
\quad "action": "get\_customer\_info",\\
\quad "parameters": \{\\
\quad\quad "NAME": "Alice Smith"\\
\quad \}\\
\}\\
</tool\_call>
\end{tcolorbox}
\end{minipage}%
\hfill
\begin{minipage}[t]{0.48\linewidth}
\noindent\textbf{\textcolor{red!70!black}{\texttimes}\ Failing Generation} \hrulefill

\begin{tcolorbox}[
  colback=red!3, colframe=red!50!black, boxrule=0.3pt,
  arc=1pt, left=3pt, right=3pt, top=3pt, bottom=3pt
]
\ttfamily\footnotesize
<tool\_call>\\
\{\\
\quad "action": "get\_customer\_info",\\
\quad "name": "Alice Smith"\\
\quad \textcolor{red!70!black}{\textrm{\scriptsize$\uparrow$ flat structure, lowercase key}}\\
\}\\
</tool\_call>
\end{tcolorbox}
\end{minipage}

\medskip

\begin{tcolorbox}[
  colback=blue!3, colframe=blue!30, boxrule=0.3pt,
  arc=1pt, left=4pt, right=4pt, top=3pt, bottom=3pt
]
\footnotesize
\textbf{Key insight:} The model correctly identifies the action and the semantic parameter (\texttt{"Alice Smith"}) in both cases. The failure is purely \emph{structural}, the decoder regresses to a standard flat JSON distribution, ignoring the non-standard nesting and uppercase constraints.
\end{tcolorbox}

\end{tcolorbox}

\caption{A tool-calling example illustrating representational load on the decoder. The model preserves semantic correctness but fails to comply with a proprietary schema that deviates from standard training distributions. The passing output respects the required \texttt{parameters} nesting and uppercase keys, the failing output falls back to a common flat JSON pattern.}
\label{fig:motiv_tool_call}
\end{figure}

LLMs are trained on massive corpora of standard programming patterns and common data formats, internalizing strong priors over these distributions. This internalization differs across models depending on their training data and size. When a task requires outputs that deviate from these familiar patterns—such as proprietary tool-calling schemas, non-standard nesting conventions, or domain-specific field naming rules—the model faces an additional \emph{representational load} beyond the core reasoning required to solve the task \cite{lu2025learning}. This burden falls disproportionately on the \emph{decoder} side of generation. Unlike an encoder, which processes the full input sequence and can leverage a global view of constraints, the autoregressive decoder must honour structural requirements \emph{token by token}, maintaining an implicit state machine that tracks nested brackets, required field names, and syntactic rules as it generates each successive token. This local, sequential nature makes the decoder inherently fragile when the target structure deviates from its training distribution: the model may correctly identify the semantic solution but fail to serialize it into the required format. We quantify this phenomenon across various models and datasets in our study.

Figure~\ref{fig:motiv_tool_call}  illustrates this concretely. The prompt requires the model to follow a strict, slightly non-standard schema: arguments must be nested inside a \texttt{parameters} object, and all keys must be uppercase. While a frontier model correctly identifies the semantic goal—calling \texttt{get\_customer\_info} with the argument \texttt{"Alice Smith"}—the failing generation regresses to a standard, flat JSON structure commonly seen in training data, ignoring the non-standard nesting and capitalization constraints. The model has not lost its understanding of the task; the failure lies entirely in the decoder's inability to manage the representational load of unfamiliar structural constraints during sequential token generation.

These observations naturally raise a question: if the structural bottleneck is concentrated on the decoder side, can targeted interventions at the decoding level alleviate it? The autoregressive decoder, constrained to commit to each token irrevocably before seeing what follows, lacks the global view that would allow it to plan ahead for complex structural requirements \cite{xiong2025unveiling}. This is precisely where techniques such as grammar-constrained decoding, regex-based validation, and template-driven token matching become relevant—they aim to externally impose the structural discipline that the decoder cannot reliably maintain on its own \cite{beurer2024guiding,park2025flexible}. However, the effectiveness of such techniques is far from settled. A critical open question is whether structural failures are primarily a matter of surface-level syntax (misplaced brackets, wrong delimiters) that can be mechanically corrected, or whether they reflect deeper issues—such as the model omitting required fields or misunderstanding schema semantics—that persist even when syntax is perfectly enforced. Moreover, the degree to which the required output format \emph{deviates} from the model's training distribution appears to be a key factor: as illustrated in Figure~\ref{fig:motiv_tool_call}, structural compliance degrades sharply with increasing distributional deviation, even as semantic accuracy remains stable. This suggests that the decoder's fragility is not uniform but may be modulated by how familiar the target structure is to the model \cite{mccoy2023embers}. Understanding this interplay between distributional familiarity, surface syntax, and deeper structural correctness is essential for designing effective mitigation strategies and for setting realistic expectations about what current tools can and cannot guarantee.

This paper is guided by the following research questions:
\begin{itemize}
\item \textbf{RQ1:} How well the LLMs perform and how frequently do structural and syntactic errors occur in LLM-generated outputs for software engineering tasks, and how can these errors be systematically categorized?


\item \textbf{RQ2:} What insights can a case study of structural output failures provide regarding error patterns and their practical impact?

\item \textbf{RQ3:} To what extent do existing structure-enforcing techniques, including grammar-constrained decoding, regex-based validation, and strict template-based generation (TTMG), reduce syntax errors and what types of errors remain after syntax control structure-enforcing tools?
\end{itemize}

To investigate the challenges of structured output generation in software engineering, we begin with a systematic evaluation of leading LLMs across a suite of representative SE tasks. Our large-scale error analysis reveals that structural and syntax-related failures—such as format violations, schema mismatches, or malformed syntax—account for a significant share of incorrect outputs, often surpassing logic or semantic errors. This finding underscores that structure adherence remains a critical bottleneck for deploying LLMs in high-reliability SE applications.

To address this issue, we examine a range of structure-enforcing techniques, including grammar-constrained decoding, regex-based filtering, and prompt engineering. While these methods help reduce syntax violations, they deliver only modest improvements in overall task success. We identify key limitations in existing solutions, including the lack of reliable grammars for many formats, runtime overhead, and limited expressiveness for enforcing fine-grained structure. In response, we introduce a novel template-driven generation approach that integrates precise token matching with dynamic mode switching between fixed-format scaffolding and free-form generation. This lightweight method guarantees structural correctness without requiring full grammars or incurring significant computational cost. 
Our implementation achieves improved format compliance across multiple SE benchmarks and demonstrates consistent gains in output fidelity, speed, and end-to-end task accuracy. 
\paragraph{Key Contributions:} In summary, our contributions are:

\begin{itemize}
\item \textbf{Systematic Evaluation of LLM Structured Outputs}: We conduct a large-scale analysis across multiple software engineering tasks, identifying structural and syntactic failures as a major source of unreliability.

\item \textbf{Error Taxonomy and Case Study}: We provide a detailed classification of structural output errors and illustrate their real-world implications through a case study of representative outputs.

\item \textbf{Benchmarking of Structure-Enforcing Techniques}: We compare grammar-constrained decoding, regex-based validation, and template-based generation (TTMG) to understand their effects on syntax, structure, and overall task accuracy.

\item \textbf{Isolation of Error Sources}: By applying TTMG as a strict control, we investigate whether eliminating syntax errors fully resolves failures, or whether deeper structural and semantic issues persist.
\end{itemize}
To facilitate replication, we have made all artifacts, software, and experiment results available\footnote{https://figshare.com/s/7d63a114a63fdf081067}.

\section{Background and Challenges}

\subsection{Structural-aware Tasks}
LLMs on generating structured outputs such as code, database queries, or JSON data remains a significant challenge~\cite{lu2025learning}. We compare the five most downloaded general-purpose LLMs against two domain-specific LLMs that have good performance in each domain, fine-tuned for structure-aware tasks like code generation or bug fixing \footnote{\url{https://huggingface.co/models?pipeline_tag=text-generation&sort=downloads}}. 
All models used in this study are summarized in Table~\ref{tab:models} and are evaluated in a zero-shot setting across four representative structured-output tasks, each aligned with a standard benchmark dataset:

\begin{table}[htbp]
\caption{Model Selected for Evaluation}
\label{tab:models}
\centering
\resizebox{0.6\columnwidth}{!}{%
\begin{tabular}{lll}
\hline
Type & Model Name         & Size      \\ \hline
\multirow{7}{*}{\begin{tabular}[c]{@{}l@{}}General\\ Purpose\\ LLMs\end{tabular}} & LLaMA-3.1-8B        & 8B     \\
     & Qwen-2.5-7b-it     & 7B       \\
     & Gemma2-9b-it       & 9B        \\
     & Qwen3-30B-A3B-IT & 30B        \\
      & Mixtral-8x7B-Instruct-v0.1 & 56B     \\
     & LLaMA-3.1-70B GGUF & 70B        \\
     & GPT-4.1-mini       & -         \\ \hline
\multirow{2}{*}{\begin{tabular}[c]{@{}l@{}}Domain\\ Specific\end{tabular}}        & Qwen2.5-Coder-7B-it & 7B  \\
     & 
Seed-Coder-8B-it     & 8B    \\ \hline
\end{tabular}%
}
\vspace{-1em}
\end{table}

\begin{itemize}
    \item \textbf{Text-to-Code Generation}: using the BigCodeBench dataset for natural-language to Python coding tasks~\cite{zhuo2024bigcodebench}.
    \item \textbf{Text-to-SQL Query Generation}: using the Spider benchmark for Text-to-SQL translation~\cite{yu2018spider}.
    \item \textbf{Text-to-JSON Output}: using tasks from the CallNavi dataset that require producing a JSON-formatted answer given an unstructured description~\cite{song2025callnavi}.
    \item \textbf{Function Calling}: using the Berkeley Function-Calling Leaderboard (BFCL) evaluation to generate function call expressions (API calls) in response to user queries~\cite{berkeley-function-calling-leaderboard}.
\end{itemize}
When we evaluate programming tasks (Python code and SQL), we employ the standard pass@1 accuracy metric, essentially the percentage of prompts for which the model’s first attempt yields a correct solution. This strict metric reflects real-use scenarios where only one output is returned. For the JSON and function-call tasks, we adopt AST-based evaluation metrics inspired by BFCL\cite{patil2024gorilla}.

\subsection{Error Types in Structured Outputs}
To understand common failure modes in structured-output generation, prior work and practical experience suggest that errors typically fall into three categories: syntax errors, structural errors, and value errors. This taxonomy provides a useful lens for characterizing the challenges faced by LLMs in producing well-formed and semantically correct outputs.

\begin{itemize}
    \item \textbf{Syntax Errors}: Cases where the model’s outputs are invalid or unparsable due to format errors (e.g. compile error, or JSON syntax mismatch). Such errors indicate the model did not follow the required syntax strictly.
    \item \textbf{Structural Errors}: Outputs that are syntactically valid but do not conform to the required structure or schema. For example, a JSON object may be well-formed but missing required fields, or a function call may violate the expected signature format. We categorize such errors as structural only when they break format constraints. If the structure is correct but semantically inappropriate (e.g., a task-irrelevant API call), it is treated as a value error.
    \item \textbf{Value Errors:} Outputs that are format-compliant (no syntax or structural issues) but semantically incorrect. In this case, the model followed the syntax and structure correctly, yet produced wrong values or logic – for example, a Python solution that runs but yields an incorrect answer, or a function call with the right structure but incorrect arguments for the user’s query.
\end{itemize}



\subsection{Pre and Post Inference Approaches}
As illustrated in Figure~\ref{fig:tools}, structure-control approaches can be broadly categorized into three domains: prompt-level methods, inference-level techniques, and post-processing or validation strategies. First, let us quickly review \textbf{pre-} (prompt) and \textbf{post-inference} approaches in this subsection.

\paragraph{Prompting and Format Instructions}, A straightforward approach is to craft prompts that explicitly instruct the LLM to produce a specific format or schema. Several studies benchmarked zero-shot and few-shot prompting methods for JSON/XML outputs. For example,  StructuredRAG, a suite of six tasks to test LLMs’ ability to follow JSON format instructions~\cite{shorten2024structuredrag}. They compared prompting strategies such as f-string templating and “follow-the-format” instructions, finding that while models can often comply, reliability is far from perfect. On average about 82.5\% of outputs were format-compliant in their experiments, but success ranged from 0\% on the hardest task to 100 on the simplest. Complex outputs (e.g. nested lists or composite objects) caused frequent errors. These results highlight that prompt engineering alone yields inconsistent structured output.

Best practices to improve prompting success include providing example outputs, explicitly showing the JSON/YAML syntax in the prompt, and requesting the model to only produce the structured portion (sometimes by wrapping the format in markdown code fences). Prompt-based methods are simple and model-agnostic, but as several works noted, even GPT-4 can occasionally violate format requirements under prompt-alone strategies. Therefore, while prompting is the first line of defense for structured generation, it often needs to be augmented with additional constraints or post-processing for high-stakes applications\cite{liu2024we}. When we conduct our benchmark, we always try to find the best prompt for our dataset's 5\% samples with these prompt tuning approaches.

\paragraph{Post-hoc Validation and Correction}, which is often used in practice: the model’s output is parsed by a JSON/XML linter or compiler, and if a format error is detected, the model is prompted to fix its answer. This loop may repeat until the output parses correctly (or a retry limit is hit). Such auto-correction frameworks have been reported to greatly reduce formatting errors by catching lapses and giving the model a chance to repair them. Amazon AWS has begun formalizing this experimental process, for instance, self-correction strategies where the LLM critiques and repairs its output, or dual-model setups where one generates and another validates~\cite{wang2025slot}. 

Overall, training-time and post-processing methods that involve schema validation show that aligning the model’s objectives with structural correctness (via RL or iterative feedback) is a promising direction to increase reliability.

\begin{figure}
    \centering
    \includegraphics[width=\linewidth]{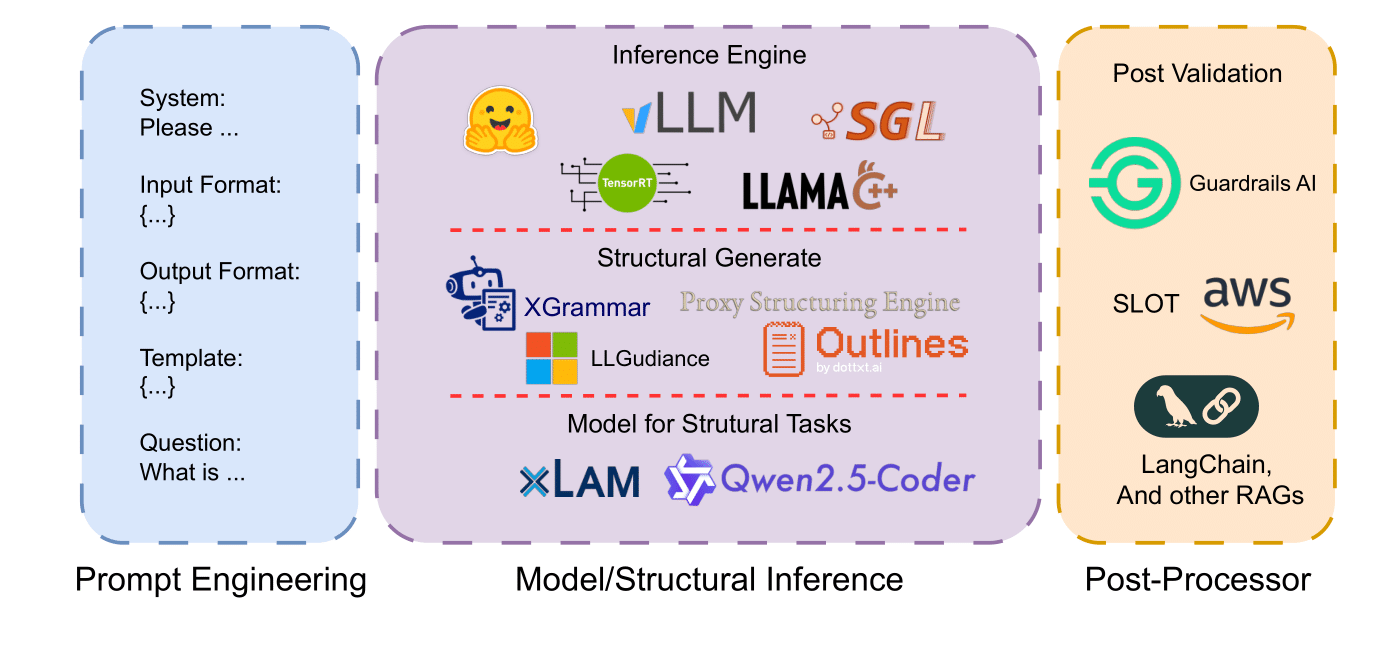}
    \caption{Current structural output solutions, divided by before inference, inference, post inference.}
    \label{fig:tools}
    \vspace{-1em}
\end{figure}

\subsection{Rule-based Structured Generation by Formats}
\begin{table*}[htbp]
\caption{Survey of Structure Output Approaches }
\centering
\label{tab:structtools}
\resizebox{\columnwidth}{!}{%
\begin{tabular}{llll}
\hline
\multicolumn{1}{c}{\textbf{Tool Name}} &
  \multicolumn{1}{c}{\textbf{Inference Engine Support}} &
  \multicolumn{1}{c}{\textbf{Rule Support}} &
  \multicolumn{1}{c}{\textbf{Output Format Support}} \\ \hline
Outlines~\cite{willard2023efficient}                     & vLLM, Transformers, SGLang  & JSON Schema, Regular Expressions                & JSON, XML, Custom Text     \\
LLGuidance~\cite{guidance2023}                     & TensorRT-LLM, llama.cpp, vLLM, SGLang  & JSON Schema, Regular Expressions, Lark-like CFG & JSON, XML, Custom Text     \\
XGrammar~\cite{dong2024xgrammar}                   & vLLM, SGLang                    & EBNF (GGML BNF), JSON schema,                                & JSON, XML, Structured Text \\
lm-format-enforcer~\cite{lm-format-enforcer}             & vLLM, Transformers              & JSON Schema, Regular Expressions                & JSON , Structured Text                      \\
Proxy Structuring Engine(PSE)~\cite{Wind_Proxy_Structuring_Engine_2025} & llama.cpp, local LLMs           & Custom FSM-based Grammar                        & JSON, XML, Custom Formats  \\
LangChain~\cite{LangChainWebsite}                      & OpenAI, HuggingFace, local LLMs & JSON Schema, Pydantic Models                    & JSON                       \\
LLM-Guard~\cite{llm-guard}               & OpenAI, HuggingFace             & JSON Schema                                     & JSON                       \\
llgtrt~\cite{llgtrt}                         & TensorRT-LLM                    & JSON Schema, Regular Expression, Lark-like CFG         & JSON, XML, Custom Formats  \\ \hline
\end{tabular}%
}
\vspace{-1em}
\end{table*}
Then, in the \textbf{inference} step, Large language models can be guided to produce outputs in strictly defined formats using structured output enforcement tools. These tools ensure that model-generated text adheres to a specified schema or grammar (e.g., valid JSON or XML), addressing the unreliability of prompt-only formatting instructions. We performed a comprehensive literature search for works published between 2022 and 2025 on structured output generation in LLMs leading to Table \ref{tab:structtools}, which summarizes prominent approaches. 

A unifying principle behind these structured decoding techniques is the use of automata (finite-state machines) to represent the allowable output language. For example, a JSON schema, RegEx or grammar can be compiled into an automaton that recognizes exactly the strings conforming to that specification. During generation, the automaton’s state is tracked in parallel with the model’s output: at any point in the parsing process there is a well-defined set of allowed next characters or tokens, given the partial output so far. If the model proposes a token that is not in this allowed set (i.e. a transition not defined by the automaton), that token is forbidden – typically by assigning it zero probability before sampling. The decoder thus only explores paths through the state-space that lead to a valid terminal state (e.g. a complete JSON object satisfying the schema). In practice, implementations construct this automaton in various ways. For simple regular formats (like regex or fixed JSON templates), a deterministic finite-state machine (FSM) can directly encode the allowed sequences. For context-free grammars that include nesting (e.g. matching braces in JSON), a pushdown automaton (an FSM with a stack) is used to handle the additional context.

\subsection{Rule-based Generation by CFG-Constrained}
Context-Free Grammar (CFG)-constrained decoding is a technique that enforces formal grammar rules during the \textbf{inference} process of LLMs. By restricting the model's output to sequences that conform to a specified CFG, this approach ensures syntactic correctness, which is crucial for tasks requiring structured outputs like JSON, SQL, or function calls.

Several frameworks have been selected from Table \ref{tab:structtools} for comparison, considering their inference engine compatibility and rule supporting:
\begin{itemize}
    \item \textbf{LLGuidance}: A low-level grammar engine capable of enforcing arbitrary CFG rules on LLM outputs. It supports grammars expressed in JSON Schema, regular expressions, or a Lark-like CFG syntax. LLGuidance evaluates grammar constraints dynamically at each token generation step, computing the set of allowed next tokens with minimal overhead (~50µs CPU per token for a 128k vocabulary). This dynamic evaluation avoids large upfront computations and allows for flexible, structured decoding. LLGuidance has been integrated into various inference engines, including vLLM, llama.cpp, and NVIDIA’s TensorRT-LLM, as well as OpenAI’s API for JSON schema enforcement.
    \item \textbf{XGrammar}: An engine designed for grammar-constrained generation using Extended Backus-Naur Form (EBNF) grammars. XGrammar compiles a given grammar (or JSON schema/Pydantic model) into an optimized automaton, which can be used by inference engines like vLLM for guided decoding. This precompilation enables significant speed-ups (up to 5× faster token generation under load) compared to earlier methods. However, the upfront compilation may introduce a one-time setup cost and may impose limits on schema complexity.
    \item \textbf{Outlines}: A framework that supports grammar-constrained generation by constructing deterministic automata upfront to represent allowed token sequences. It primarily supports JSON schema and regular expression constraints. Similar to XGrammar, Outlines precomputes token masks to guide the generation process, which can introduce a one-time setup cost but allows for efficient decoding thereafter.
\end{itemize}
These frameworks differ in their approaches: LLGuidance evaluates grammar constraints dynamically per token, offering flexibility without significant startup costs, while XGrammar and Outlines rely on precomputed automata, which can be more efficient during generation but require pre-compilation.

By integrating CFG-constrained decoding, these tools enhance the reliability of LLMs in producing syntactically valid structured outputs, which is essential for applications that demand strict adherence to predefined formats.

\subsection{Grammars of CFG}
\lstset{
    language=Python,
    basicstyle=\ttfamily\small,
    keywordstyle=\color{blue}\bfseries,
    commentstyle=\color{gray}\itshape,
    stringstyle=\color{red},
    backgroundcolor=\color{lightgray!20},
    numbers=left,
    numberstyle=\tiny\color{darkgray},
    frame=single,
    framesep=5pt,
    tabsize=4,
    breaklines=true,
    captionpos=b,
}

To identify publicly available Lark and GBNF grammars for ten major programming languages, we systematically searched open repositories, research datasets, and community-maintained grammar libraries. The search focused on GitHub, Hugging Face datasets, and the GGML community projects. We collected all grammars explicitly written in Lark EBNF or GBNF (GGML EBNF) syntax, regardless of completeness or compatibility. Each grammar was manually inspected to verify its intended language coverage (full or partial) and testability using standard parsing interfaces. Table \ref{tab:gm} summarizes the findings, including the resource URLs and tool compatibility across both grammar systems.

\begin{table*}[h!]
\centering
\scriptsize
\caption{Publicly available Lark and GBNF grammars for 10 programming languages(Until March 2026). Full = covers entire syntax; Limited = partial subset. Tool Compatibility indicates known usability in grammar-constrained decoding frameworks.}
\label{tab:gm}
\resizebox{\columnwidth}{!}{
\begin{tabular}{|l|l|l|l|l|l|l|}
\hline
\textbf{Language} & \textbf{Lark Grammar} & \textbf{Lark URL} & \textbf{Compatibility} & \textbf{GBNF Grammar} & \textbf{GBNF URL} & \textbf{Compatibility} \\
\hline
Python & Limited & \href{https://github.com/ligurio/lark-grammars}{ligurio/lark-grammars (python3.lark)} & Compatible & None & -- & N/A \\
\hline
C++ & None & -- & N/A & None & -- & N/A \\
\hline
C & Limited & \href{https://github.com/kaby76/Domemtech.TrashBase}{kaby76/Domemtech.TrashBase (C2.lark)} & Fail Tests  & Limited & \href{https://github.com/ggerganov/llama.cpp/blob/master/grammars/c.gbnf}{llama.cpp (c.gbnf)} & Fail Tests \\
\hline
Java & Limited & \href{https://github.com/structuredllm/syncode}{structuredllm/syncode} & Fail Tests & None & -- & N/A \\
\hline
C\# & None & -- & N/A & None & -- & N/A \\
\hline
JavaScript & None & -- & N/A & None & -- & N/A \\
\hline
Go & Limited & \href{https://github.com/structuredllm/syncode}{structuredllm/syncode} & Fail Tests & None & -- & N/A \\
\hline
Visual Basic & None & -- & N/A & None & -- & N/A \\
\hline
Delphi & None & -- & N/A & None & -- & N/A \\
\hline
SQL & Full & \href{https://github.com/structuredllm/syncode}{structuredllm/syncode} & Compatible & None & -- & N/A \\
\hline
\end{tabular}
}
\end{table*}

To evaluate grammar usability, we designed a minimal parsing benchmark using both \texttt{Lark} and \texttt{XGrammar} toolchains. The benchmark employed a simple procedural program containing \textbf{sequential execution}, \textbf{looping}, and \textbf{conditional branching} constructs. An example of the test code is shown below:
\lstset{language=Python}
\begin{lstlisting}[language=Python, caption={Example string-processing test code used for grammar validation.}]
SET x TO 0
FOR i FROM 0 TO 4 DO
    IF (i MOD 2 = 0) THEN
        x = x + i
    ELSE
        x = x - i
    END IF
END FOR
PRINT x
\end{lstlisting}

Each grammar was loaded and compiled using the corresponding parsing backend.

The evaluation results show that only the \textbf{Python (Lark)} and \textbf{SQL (Lark)} grammars successfully parsed the test programs. All other grammars—either incomplete or inconsistent—failed to process even basic control-flow constructs (loops or branches). This finding indicates that, despite partial community efforts, valid and usable grammars remain scarce for most major programming languages, especially in \textbf{GBNF format}. Consequently, developing standardized and well-tested grammar representations is an open challenge for enabling robust grammar-constrained code generation.

\section{Empirical Analysis and Experiments}
\subsection{Experiments for no-constrained LLM outputs}
\begin{table}[htbp]
\caption{Result for Zero-shot General and Domain Specific Models}
\centering
\label{tab:nonstruct-result}
\resizebox{\columnwidth}{!}{%
\begin{tabular}{llrrrrr}
\hline
\multirow{2}{*}{Models} &
  \multirow{2}{*}{Tasks} &
  \multicolumn{1}{l}{Overall Score} &
  \multicolumn{3}{c}{Error Type} &
  \multicolumn{1}{l}{} \\ \cline{3-7} 
 &
   &
  \multicolumn{1}{l}{\begin{tabular}[c]{@{}l@{}}Pass@1/\\ Exact Match\end{tabular}} &
  \multicolumn{1}{l}{Syntax} &
  \multicolumn{1}{l}{Structural/Runtime} &
  \multicolumn{1}{l}{Value} &
  \multicolumn{1}{l}{Total Cases} \\ \hline
\multirow{4}{*}{LLaMA-3.1-8B}           & BigCodeBench & 0.2866 & 33   & 40  & 726 & 1140 \\
                                        & Spider       & 0.6767 & 42   & 0   & 173 & 665  \\
                                        & CallNavi     & 0.5158 & 12   & 309 & 41  & 729  \\
                                        & Bfcl v2      & 0.1759 & 1833 & 119 & 85  & 2501 \\ \hline
\multirow{4}{*}{Qwen-2.5-7b-it}         & BigCodeBench & 0.3427 & 14   & 22  & 713 & 1140 \\
                                        & Spider       & 0.7278 & 2    & 29  & 152 & 665  \\
                                        & CallNavi     & 0.5583 & 5    & 303 & 21  & 729  \\
                                        & Bfcl v2      & 0.5306 & 79   & 586 & 460 & 2501 \\ \hline
\multirow{4}{*}{Gemma2-9b-it}           & BigCodeBench & 0.1262 & 399  & 35  & 562 & 1140 \\
                                        & Spider       & 0.5308 & 168  & 0   & 144 & 665  \\
                                        & CallNavi     & 0.0988 & 586  & 61  & 10  & 729  \\
                                        & Bfcl v2      & 0.0472 & 2203 & 69  & 32  & 2501 \\ \hline
\multirow{4}{*}{Qwen3-30B-A3B-IT}     & BigCodeBench & 0.4473 & 0    &  24 & 606 & 1140 \\
                                        & Spider       & 0.7687 & 2   &  4  &147  & 665  \\
                                        & CallNavi     & 0.5665 &  5  &  290 &  22 & 729  \\
                                        & Bfcl v2      & 0.5033 & 134  & 622 & 486 & 2501 \\ \hline
\multirow{4}{*}{Mixtral-8x7B-IT}     & BigCodeBench & 0.1807 & 303    &  36 & 595 & 1140 \\
                                        & Spider       & 0.6601 & 23       & 30  & 173 &665  \\
                                        & CallNavi     & 0.4128 &  73  &  341 &  14 & 729  \\
                                        & Bfcl v2      & 0.4086 & 669  & 395 & 415 & 2501 \\ \hline
\multirow{4}{*}{LLaMA-3.1-70B GGUF}     & BigCodeBench & 0.4224 & 1    & 43  & 614 & 1140 \\
                                        & Spider       & 0.7553 & 10   & 0   & 152 & 665  \\
                                        & CallNavi     & 0.5556 & 19   & 284 & 24  & 729  \\
                                        & Bfcl v2      & 0.4934 & 297  & 475 & 474 & 2501 \\ \hline

\multirow{4}{*}{GPT-4.1-mini}           & BigCodeBench & 0.4873 & 1    & 33  & 550 & 1140 \\
                                        & Spider       & 0.7654 & 0    & 7   & 149 & 665  \\
                                        & CallNavi     & 0.5748 & 1    & 297 & 15  & 729  \\
                                        & Bfcl v2      & 0.4820 & 151  & 636 & 488 & 2501 \\ \hline
\multirow{4}{*}{Qwen-2.5-Coder-7B-it}   & BigCodeBench & 0.3672 & 18   & 7   & 684 & 1140 \\
                                        & Spider       & 0.7838 & 1    & 14  & 128 & 665  \\
                                        & CallNavi     & 0.3923 & 251  & 183 & 11  & 729  \\
                                        & Bfcl v2      & 0.5042 & 113  & 569 & 493 & 2501 \\ \hline
\multirow{4}{*}{Seed-Coder-8B-Instruct} & BigCodeBench & 0.4400 & 13   & 5   & 609 & 1140 \\
                                        & Spider       & 0.6857 & 3    & 16  & 193 & 665  \\
                                        & CallNavi     & 0.3580 & 267  & 185 & 25  & 729  \\
                                        & Bfcl v2      & 0.4456 & 286  & 613 & 417 & 2501 \\ \hline
\end{tabular}%
}
\vspace{-1em}
\end{table}
We evaluate all models under three decoding backends—vLLM, SGLang, and HuggingFace Transformers—and report the highest performance achieved for each model. All experiments run on two RTX 5000 Ada GPUs, with models configured using their default parameters. To ensure fair comparisons, we conduct a small batch of prompt tuning trials for each task using GPT-4.1, following best practices from structured prompt engineering guides, and apply the best-found prompts uniformly across models.

We use a prompt to classify error types and extract keywords: ``What error type and keyword from the following \{Python/SQL\} error message is? Give me an error class and up to 3 keywords related to the message.''. For example, exception types such as 'SyntaxError' or keywords such as 'missing field' are generated and facilitate a statistical analysis of error patterns. This helps identify which error types are most prevalent for each model and task.  

As shown in Table \ref{tab:nonstruct-result} and Fig. \ref{fig:piechart}, unconstrained decoding yields substantial structure-related errors:
\begin{itemize}
    \item CallNavi and BFCL v2 show the highest failure rates: syntax and structural errors together account for around \~80.7\% of all failures, with LLaMA-3.1-8B producing 1833 syntax errors in BFCL v2.
    \item BigCodeBench and Spider (Coding tasks) are less affected by format issues (less than 20\% of failures), but still exhibit invalid outputs that impact usability.
    \item Larger models like GPT-4.1-mini and LLaMA-3.1-70B perform better but remain imperfect, suggesting that scale alone is insufficient for format fidelity. Notably, the MoE model Mixtral-8x7B-IT produces 669 syntax errors on BFCL~v2 and scores only 0.1807 on BigCodeBench, underperforming even smaller dense models.

    \item Common error types (Fig. \ref{fig:piechart}), including SyntaxError, no such column, and TypeError, indicate frequent failures in structural and schema adherence.
\end{itemize}

Our analysis, detailed with error type distributions in Figure \ref{fig:pie}, reveals that Name \& Attribute, Type \& Value, and Key \& Index errors form the bulk of syntax and structural issues in unconstrained LLM outputs. This establishes the methodology for subsequent LLM performance comparisons and underscores that effective structure control is vital for reliable LLM use in SE, especially with strict output requirements.

\begin{figure}
    \centering

    \includegraphics[width=0.7\linewidth]{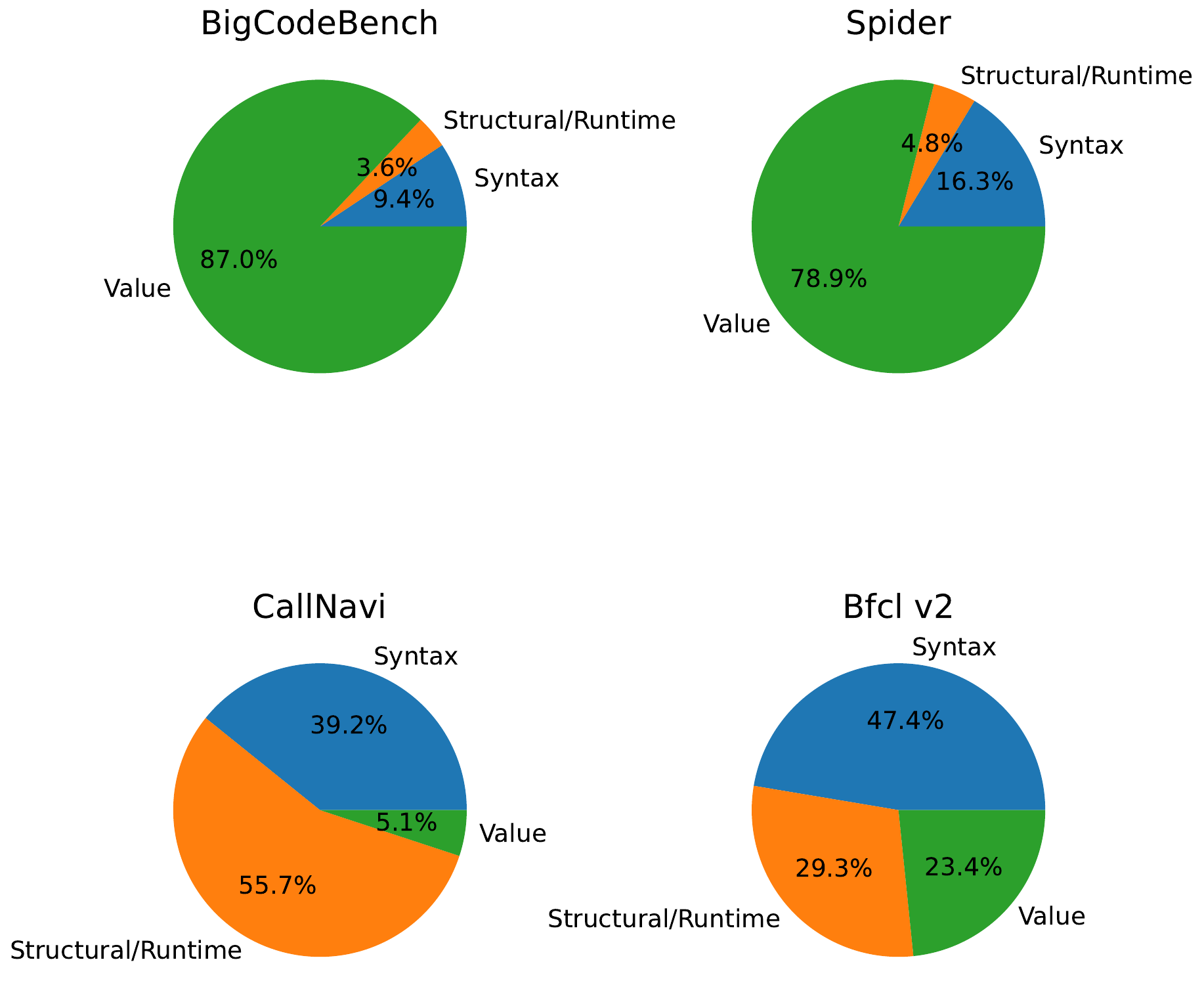}
    \caption{Pie Chart statistic of error types from each dataset from all testing results in Table \ref{tab:nonstruct-result}}
    \label{fig:piechart}
    \vspace{-1em}
\end{figure}

\begin{figure}
    \centering

    \includegraphics[width=0.7\linewidth]{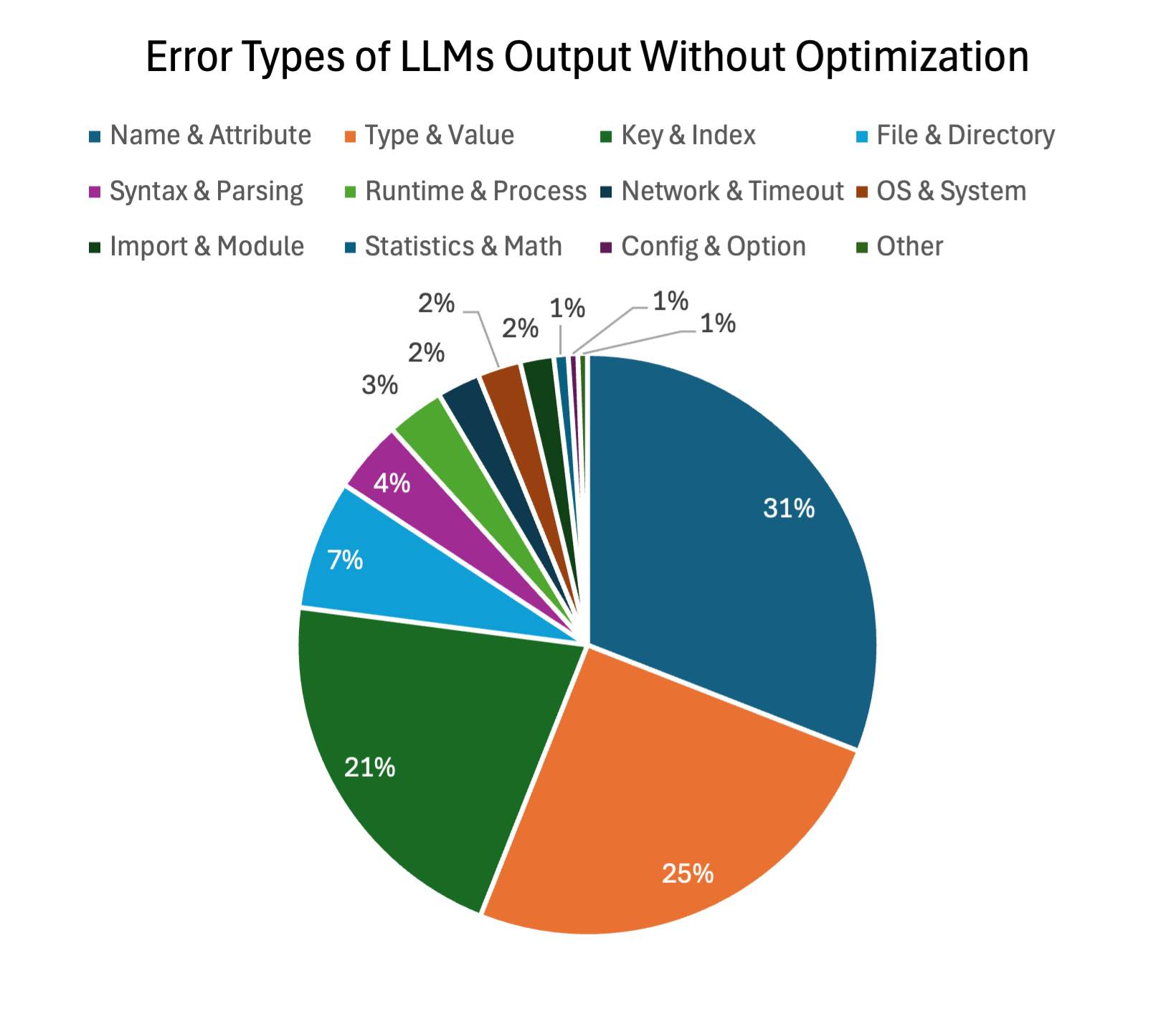}
    \caption{Pie Chart of unconstrained error type classification by GPT 4.1}
    \label{fig:pie}
    \vspace{-1em}
\end{figure}

\begin{tcolorbox}[leftrule=0mm,rightrule=0mm,toprule=0mm,bottomrule=0mm,left=0pt,right=0pt,top=0pt,bottom=0pt,title={RQ1: How frequently do structural and syntactic errors occur in LLM-generated outputs for software engineering tasks, and how do these errors affect downstream task performance?}]
    \textbf{Answer}: Structural and syntax errors are highly prevalent in state-of-the-art LLM output when performing structured software engineering tasks. These errors, such as invalid JSON, incorrect function signatures, or malformed SQL queries, represent a substantial share of all failures, sometimes exceeding logical or semantic mistakes. As a result, such errors directly compromise the usability and reliability of LLM-generated outputs, often rendering them unusable in downstream SE workflows.
\end{tcolorbox}

\subsection{Experiment for Rule-based Structured Generation}

To evaluate the impact of structural control methods with grammar/rule constraints, we tested three representative tools: Outlines, LLGuidance, and XGrammar, in three structured output tasks: JSON (CallNavi), SQL (Spider), and function calling (BFCL v2). All experiments are conducted using the SGLang framework. Unfortunately, we cannot find a bug-free Python grammar for the BigCodeBench dataset. For grammar-based decoding, we use SQL and JSON Lark grammars~\cite{faria2024upgrade}, which are widely supported in table \ref{tab:structtools}; for function calling, we apply a common RegEx pattern matching valid API invocation formats. Each experiment is repeated three times, and we report the median result. 

Table~\ref{tab:structresult} provides a detailed evaluation of several structure-enforcement tools across various large language models and structure-aware tasks, assessing exact match rates, syntax errors, structural errors, and semantic correctness via AST matching. Our findings indicate that grammar-based methods, such as Outlines and XGrammar, as well as regex-based constraints, consistently achieve substantial reductions in syntax errors but yield only modest improvements in task-level accuracy metrics like exact match and pass@1 compared to unconstrained decoding. Specifically, exact match scores on CallNavi and SQL tasks remain relatively similar across structured and unconstrained conditions, illustrating that simply ensuring syntax validity does not inherently resolve underlying semantic or logical errors. This discrepancy underscores a critical nuance: while grammar- and regex-based tools effectively eliminate superficial syntax errors—crucial for applications requiring strictly valid outputs—they typically do not rectify semantic inaccuracies or logical flaws. Thus, while structural control methods are valuable for enforcing syntactic correctness, they must be complemented by deeper semantic alignment techniques to achieve meaningful improvements in end-task performance.

\begin{table*}[htbp]
\caption{Test results for selected structure output tools. For JSON (CallNavi) and SQL (Spider) tasks, we tested CFG grammar-based tools using Lark grammar. For function calling (BFCL v2), we used the same tools but configured them with regular expressions to enforce API invocation formats. The result is in the format Syntax/Structural/Runtime Error samples. Total number see table \ref{tab:nonstruct-result}. We run experiments 3 times and take the Median.}
\label{tab:structresult}
\resizebox{\columnwidth}{!}{%
\begin{tabular}{llrrrrrrrrr}
\hline
\multirow{3}{*}{Models} &
  \multirow{3}{*}{Tools} &
  \multicolumn{3}{c|}{CallNavi} &
  \multicolumn{3}{c|}{Spider} &
  \multicolumn{3}{c}{BFCL v2} \\ \cline{3-11} 
 &
   &
  \multicolumn{1}{l}{Exact Match} &
  \multicolumn{1}{l}{Syntax Error} &
  \multicolumn{1}{l|}{Struct Error} &
  \multicolumn{1}{l}{Pass@1} &
  \multicolumn{1}{l}{Syntax Error} &
  \multicolumn{1}{l|}{Runtime Error} &
  \multicolumn{1}{l}{AST Match} &
  \multicolumn{1}{l}{Syntax Error} &
  \multicolumn{1}{l}{Struct Error} \\ \cline{3-11} 
 &
   &
  \multicolumn{6}{c|}{CFG(Lark/GBNF) Grammar} &
  \multicolumn{3}{c}{RegEx} \\ \hline
\multirow{4}{*}{LLaMA-3.1-8B} &
  Outlines &
  0.5130 &
  20 &
  305 &
  0.7113 &
  8 &
  25 &
  0.5274 &
  2 &
  468 \\
 &
  LLGuidance &
  0.4883 &
  3 &
  339 &
  0.6992 &
  3 &
  24 &
  0.5230 &
  0 &
  480 \\
 &
  XGrammar &
  0.5130 &
  7 &
  329 &
  - &
  - &
  - &
  0.5230 &
  19 &
  472 \\
 &
  \textit{N/A} &
  \textit{0.4883} &
  12 &
  309 &
  \textit{0.6767} &
  8 &
  34 &
  \textit{0.1759} &
  1833 &
  119 \\ \hline
\multirow{4}{*}{Qwen-2.5-7b-it} &
  Outlines &
  0.5624 &
  6 &
  303 &
  0.7218 &
  16 &
  21 &
  0.5354 &
  6 &
  603 \\
 &
  LLGuidance &
  0.5596 &
  4 &
  302 &
  0.7248 &
  6 &
  19 &
  0.5370 &
  6 &
  606 \\
 &
  XGrammar &
  0.5651 &
  2 &
  307 &
  - &
  - &
  - &
  0.5238 &
  8 &
  607 \\
 &
  \textit{N/A} &
  \textit{0.5583} &
  5 &
  303 &
  \textit{0.7278} &
  1 &
  14 &
  \textit{0.5042} &
  79 &
  586 \\ \hline
  
  \multirow{3}{*}{Qwen3-30B-A3B-IT} &
  Outlines &
  0.5665 &
  8 &
  291 &
  0.7702 &
  7 &
  1 &
  0.5073 &
  123 &
  622 \\
 &
  LLGuidance &
  0.5679 &
  6 &
  292 &
  0.7669 &
  1 &
  0 &
  0.5065 &
  131 &
  618 \\
 &
  \textit{N/A} &
  \textit{0.5665} &
  5 &
  290 &
  \textit{0.7687} &
  2 &
  4 &
  \textit{0.5033} &
  134 &
  622 \\ \hline

  \multirow{3}{*}{Mixtral-8x7B-IT} &
  Outlines &
  0.4183 &
  68 &
  342 &
  0.6511 &
22 &
  28 &
  0.5277 &
  36 &
  503 \\
 &
  LLGuidance &
  0.4128 &
  67 &
  344 &
  0.6556 &
  20 &
  35 &
  0.2363 & 78
   & 408
   \\
 &
  \textit{N/A} &
  \textit{0.4128} &
  73 &
  341 &
  \textit{0.6601} &
  23 &
  30 &
  \textit{0.4086} &
  669 &
  395 \\ \hline
\multirow{3}{*}{Seed-Coder-8B-Instruct} &
  Outlines &
  0.5802 &
  7 &
  274 &
  0.7098 &
  10 &
  5 &
  0.4886 &
  1 &
  678 \\
 &
  XGrammar &
  0.5761 &
  1 &
  286 &
  - &
  - &
  - &
  0.4794 &
  42 &
  666 \\
 &
  \textit{N/A} &
  \textit{0.3580} &
  267 &
  185 &
  \textit{0.6857} &
  3 &
  16 &
  \textit{0.4456} &
  286 &
  613 \\ \hline
\multirow{4}{*}{Qwen-2.5-Coder-7B-it} &
  Outlines &
  0.5583 &
  11 &
  297 &
  0.7820 &
  2 &
  15 &
  0.4962 &
  5 &
  586 \\
 &
  LLGuidance &
  0.5542 &
  7 &
  301 &
  0.7789 &
  3 &
  13 &
  0.4926 &
  7 &
  618 \\
 &
  XGrammar &
  0.5665 &
  1 &
  305 &
  - &
  - &
  - &
  0.4946 &
  19 &
  575 \\
 &
  \textit{N/A} &
  \textit{0.3923} &
  251 &
  183 &
  \textit{0.7838} &
  1 &
  14 &
  \textit{0.5042} &
  113 &
  569 \\ \hline
\end{tabular}%
}
\end{table*}
\paragraph{Mixture-of-Experts and the Limits of Decoder-Side Constraints.}
An unexpected finding emerges when we examine the two Mixture-of-Experts (MoE) models in Table~\ref{tab:structresult}: Qwen3-30B-A3B and Mixtral-8x7B-IT. In unconstrained settings, Qwen3-30B-A3B achieves competitive performance, 0.5665 exact match on CallNavi, 0.7687 pass@1 on Spider, and 0.5033 AST match on BFCL~v2, placing it on par with GPT-4.1-mini despite activating only 3B parameters per token. However, unlike the dense models in our evaluation, both MoE models show \emph{virtually no improvement} when structure-enforcing tools are applied. For Qwen3-30B-A3B, Outlines leaves CallNavi exact match unchanged at 0.5665 and structural errors move from 290 to 291, which we consider to be within stochastic noise. On BFCL~v2, syntax errors drop only marginally from 134 to 123 with Outlines (contrast LLaMA-3.1-8B, where the same tool reduces syntax errors from 1,833 to 2), and structural errors remain static at 622.

Mixtral-8x7B-IT reinforces this pattern more starkly. Its unconstrained BFCL~v2 baseline already shows 669 syntax errors and an AST match of only 0.4086 (Table~\ref{tab:nonstruct-result}). Under Outlines, syntax errors drop to 36 but AST match rises only to 0.5277; under LLGuidance, AST match \emph{collapses} to 0.2363, substantially \emph{worse} than the unconstrained baseline, while structural errors remain high at 408. On CallNavi, both tools leave performance effectively unchanged at $\sim$0.41, with an exact match of $\sim$340 structural errors.

This behavior is architecturally revealing. In a dense transformer, every parameter participates in every forward pass, so decoder-side logit constraints directly modulate the full model's token distribution. In an MoE architecture, only a sparse subset of experts is activated per token, selected by a learned routing mechanism in the intermediate layers. The structure-enforcing tools we evaluate operate exclusively at the logit level, masking or re-weighting token probabilities \emph{after} expert routing has already occurred. If routing has committed the active experts to a structurally non-compliant trajectory, post-hoc logit masking can suppress individual invalid tokens but cannot redirect the upstream expert selection. The result, visible in both Qwen3 and especially Mixtral, is that surface-level syntax may improve slightly while deeper structural patterns persist or, as in Mixtral's LLGuidance case, the constraint actively disrupts the model's generation, producing worse overall outputs.

\subsection{Stability Analysis of Rule-based Structured Generation}

This section differs from the previous experiment (Table~\ref{tab:structresult}) by shifting focus from task performance (e.g., accuracy and execution correctness) to output stability—that is, the consistency and fluency of generated outputs under structural constraints. Here we assess how structure-enforcing tools influence variability and consistency in generation quality, regardless of correctness.

We evaluate stability using three metrics on the Spider dataset: pass@1, pass@5, and the average BLEU score of the first output. Table~\ref{tab:stab} presents results for two LLMs—LLaMA-3.1-8B and Qwen-2.5-7B-it—each tested under three settings: with Outlines, LLGuidance, and a no-grammar baseline.

Across both models, we observe that structure-enforcing methods have minimal impact on output stability. For example, LLaMA-3.1-8B experiences only a slight decrease in pass@1 when using LLGuidance (-0.007 relative to baseline), and BLEU scores remain largely unchanged across all configurations. A similar pattern holds for Qwen-2.5-7B-it, where both 2 tools yield nearly identical results to the baseline.

These findings indicate that adding syntactic constraints does not significantly degrade the fluency or diversity of outputs, reinforcing the practicality of tools like LLGuidance and Outlines in applications where structural validity is essential but generation stability must be preserved.

\begin{table}[htbp]
\caption{Stability Test Result for spider dataset, including pass@1(median result), pass@5 and average BLEU score from the first output.}
\label{tab:stab}
\resizebox{\columnwidth}{!}{%
\begin{tabular}{llrrr}
\hline
\multirow{2}{*}{Models} & \multirow{2}{*}{Tools} & \multicolumn{3}{c}{Spider@Lark Grammar}                                                 \\ \cline{3-5} 
                        &                        & \multicolumn{1}{l}{pass@1} & \multicolumn{1}{l}{pass@5} & \multicolumn{1}{l}{Avg. BLEU} \\ \hline
\multirow{3}{*}{LLaMA-3.1-8B}   & Outlines   & 0.6937 & 0.8363 & 0.5176 \\
                                & LLGuidance & 0.6802 & 0.8438 & 0.4870 \\
                                & No Grammar & 0.6727 & 0.8243 & 0.4943 \\ \hline
\multirow{3}{*}{Qwen-2.5-7b-it} & Outlines   & 0.7267 & 0.7868 & 0.7641 \\
                                & LLGuidance & 0.7252 & 0.7868 & 0.7681 \\
                                & No Grammar   & 0.7147 & 0.7883 & 0.7646 \\ \hline
\end{tabular}%
}
\vspace{-1em}
\end{table}
\subsection{Limitations in Rule-Based Structured Generation}

Grammar-constrained decoding has recently gained attention as a method to improve syntactic correctness in code generation. By guiding language models to produce outputs that strictly adhere to a formal grammar—typically defined in CFG formats—such methods aim to reduce invalid or unparsable code. However, their effectiveness in practice depends critically on the availability of high-quality, parser-compatible grammars for the target programming languages.

To assess the practical feasibility of this approach, we conducted a systematic survey of publicly available CFGs in two common formats: Lark grammars, used with the Outlines, and GGML EBNF (GBNF) grammars\footnote{\url{https://github.com/ggml-org/llama.cpp/blob/master/grammars}}, used in tools like XGrammar to define decoding constraints. We focused on the top 10 most-used programming languages according to the May 2025 TIOBE index.\footnote{\url{https://www.tiobe.com/tiobe-index/}} Our goal was to determine whether reliable, parser-ready grammars exist for these languages that could be directly used for rule-constrained decoding.

As shown in Table~\ref{tab:grammar-lan}, the results reveal a stark gap between theory and practice. Only a few of these top languages have any public Lark or GBNF grammar available. Even among those that do, most grammars are marked as Limited—they are incomplete, unavailable, or incompatible with standard Earley or LALR(1) parsers without substantial manual adaptation. For example, while some Python and SQL grammars exist (many derived from the SynCode project~\cite{ugare2024syncode}), languages like C++, JavaScript, Visual Basic, and Delphi have no suitable Lark or GBNF grammars at all. GBNF coverage is particularly poor, with only partial experimental efforts available for C.

While our initial grammar evaluation (Section 2.6) showed that a simple Python Lark grammar could parse basic control flows, applying this grammar to the complex, real-world Python solutions generated in the BigCodeBench dataset resulted in severe toolchain failures. Specifically, the Lark grammars introduced parser-level bugs and crashed the decoding process in both Outlines and LLGuidance due to the inability to reliably handle Python's indentation and complex nested structures. Therefore, BigCodeBench was excluded from Table \ref{tab:structresult}. This failure is a critical empirical finding: \textbf{it demonstrates that current CFG-constrained decoding tools are not yet mature enough to handle complex, production-level Python code generation.}

This scarcity of robust, reusable grammars, particularly in the GBNF format required by tools like Outlines and LLGuidance, poses a major bottleneck for scaling CFG-based code generation across languages. Our results across all evaluated models (LLaMA-3.1-8B, Qwen-2.5-7B-it, and Seed-Coder-8B-Instruct) show that while grammar and regex-based constraints improve syntactic validity, they fall short of guaranteeing full compliance with complex task schemas. This underscores the need not only for community-driven grammar standardization, but also for more advanced decoding strategies, such as template-based or semantic-aware methods, to bridge the gap between structural correctness and true task success in structured generation.

\begin{tcolorbox}[leftrule=0mm,rightrule=0mm,toprule=0mm,bottomrule=0mm,left=0pt,right=0pt,top=0pt,bottom=0pt,title={RQ2: To what extent do existing structure-enforcing techniques, such as grammar/rule-constrained decoding, influence the validity, effectiveness, and stability of LLM outputs in structured SE tasks?}]
    \textbf{Answer:} Rule-based Structure-enforcing tools, including grammar and regex decoding, are effective at reducing syntax and structural errors. These methods can lower the rate of format violations from hundreds or thousands of cases to near zero. However, their impact on overall task accuracy—such as exact-match or pass@1 metrics—remains limited, with only minor improvements observed. While these tools greatly enhance format reliability, they do not consistently resolve underlying logic or value errors.
\end{tcolorbox}

\begin{table}[htbp]
\caption{Current Languages Supports for Top-10 Programming Languages.}
\centering
\label{tab:grammar-lan}
\resizebox{0.8\columnwidth}{!}{%
\begin{tabular}{lll}
\hline
Languages                                          & Lark Support & GGML EBNF(GBNF) \\ \hline
Python                                             & Limited      &      No           \\
C++                                                &  No            &           No      \\
C                                                  &    Limited          &     Limited\tablefootnote{\url{https://github.com/ggml-org/llama.cpp/blob/master/grammars/c.gbnf}}            \\
Java                                               &      Yes        &      No           \\
C\#                                                 &       No       &       No          \\
JavaScript                                         &       No       &        No         \\
Go                                                 &     Limited         &    No             \\
Visual Basic                                       &      	No        &        	No         \\
Delphi                                             &    No          &       No          \\
 SQL & Yes\tablefootnote{Refined by us.}          &         No        \\ \hline
\end{tabular}%
}
\vspace{-1em}
\end{table}

\section{A Comparison Study with Template Filling}
\subsection{Task Definition}

When the expected output format is well-defined—such as a JSON schema or a function-calling template—the task moves beyond general grammar-constrained generation and becomes suitable for \textbf{template-driven structured output}. In this setting, the input to the model consists of (1) a natural language instruction or query, and (2) a rigid, predefined template outlining the exact structure and fields to be generated. For example, given an instruction such as \textit{"Provide the weather forecast for Paris tomorrow."}, paired with a template:

\begin{small}
\begin{verbatim}
{"city": "<city_name>", 
"date": "<forecast_date>",
"temperature": "<temperature_celsius>", 
"condition": "<weather_condition>"}
\end{verbatim}
\end{small}
the structured output must exactly follow the template:
\begin{small}
\begin{verbatim}
{"city": "Paris", 
"date": "2024-05-26",
"temperature": "22", 
"condition": "Partly cloudy"}
\end{verbatim}
\end{small}

In contrast, \textbf{grammar-based structured output} methods rely on defining general rules or schemas (e.g., JSON Schema, regular expressions) without prescribing exact output patterns. Such grammars restrict model outputs within structural boundaries, allowing flexibility in field ordering, optional fields, and formatting details. This flexibility often produces diverse outputs but may also increase syntax or structural errors due to the broader search space.

Template-driven approaches instead enforce strict structural conformity, which virtually eliminates syntax violations. However, this rigidity does not guarantee semantic accuracy: placeholders may still be filled incorrectly, leading to persistent structural or logical errors. Thus, template filling provides a useful \textbf{control setting} to isolate whether problems in LLM structured output arise solely from syntax violations or whether deeper structural and semantic challenges remain.

\subsection{Template Token Match Generation as a Control}

\begin{figure}[htbp]
    \centering
    \includegraphics[width=0.7\linewidth]{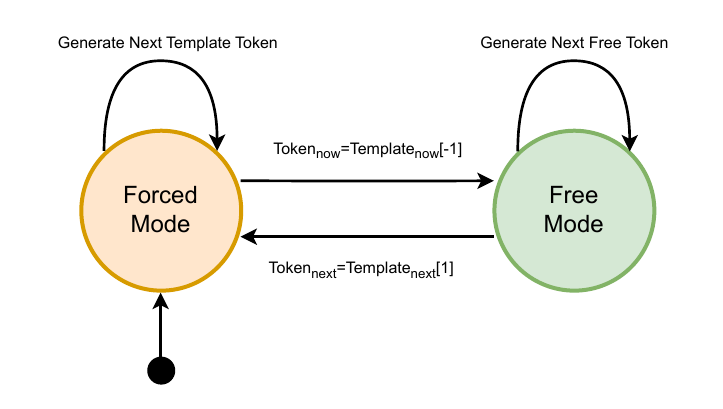}
    \caption{FSM for the token match (TTMG) logit processor}
    \vspace{-1em}
    \label{fig:fsm}
\end{figure}

Inspired by DOMINO~\cite{beurer2024guiding}, we employ a decoding strategy called \textbf{Template Token Match Generation (TTMG)}. TTMG enforces strict compliance with a given template through token matching and dynamic mode-switching: alternating between fixed-token emission and free-form placeholder filling. Unlike CFG-based methods that permit variability in structure, TTMG guarantees that outputs exactly follow a prescribed format (Figure~\ref{fig:fsm}).  Detailed description of algorithms in the appendix \ref{apd:alg}.

In this study, TTMG is \textbf{not introduced as a new solution}, but rather as a \textbf{control mechanism}. By ensuring that syntax errors are eliminated, TTMG allows us to investigate whether residual failures in LLM structured outputs are due to surface-level syntax issues or deeper structural and semantic limitations. This distinction is critical for understanding the true bottlenecks in deploying LLMs for software engineering tasks.

\subsection{Examples of TTMG}

\subsection{Comparative Experiments}

We compare template-driven generation (TTMG) with two widely used rule-based baselines: JSON Schema-constrained and RegEx-constrained decoding. Both are implemented via the XGrammar library within the SGLang generation framework. Because these approaches assume different levels of format specification, they are not directly comparable in all settings: for example, JSON Schema cannot be applied to BFCL v2 outputs, so only RegEx is used there.

We evaluate the three methods on two representative structured-output tasks—CallNavi (JSON-format) and BFCL v2 (function-calling with custom syntax)—using LLaMA-3.1-8B and Qwen-2.5-7B-it. Results are summarized in Table~\ref{tab:templated}.

Our experiments show that TTMG virtually eliminates syntax errors, as expected under strict enforcement. However, structural and logical errors persist, even when outputs are guaranteed to be syntactically correct. In comparison, grammar- and regex-based methods reduce syntax violations but still allow malformed structures and inconsistent semantics. The key insight is that syntax enforcement alone, even under the strictest conditions, does not fully resolve the broader reliability challenges of LLM structured outputs.

\begin{table*}[htbp]
\caption{Test Results for template structure generation on CallNavi and BFCL v2, N/A is the unconstrained output.}
\label{tab:templated}
\resizebox{\columnwidth}{!}{%
\begin{tabular}{ll|rrrr|rrrr}
\hline
\multirow{2}{*}{Models}         & \multirow{2}{*}{Tool} & \multicolumn{4}{l}{CallNavi} & \multicolumn{4}{l}{BFCL v2} \\ \cline{3-10} 
 &
   &
  \multicolumn{1}{l}{AST Match} &
  \multicolumn{1}{l}{Syntax Error} &
  \multicolumn{1}{l}{Structural Error} &
  \multicolumn{1}{l}{Value Error} &
  \multicolumn{1}{l}{AST Match} &
  \multicolumn{1}{l}{Syntax Error} &
  \multicolumn{1}{l}{Structural Error} &
  \multicolumn{1}{l}{Value Error} \\ \hline
\multirow{4}{*}{LLaMA-3.1-8B}   & TTMG                  & 462   & 28   & 56    & 185   & 1874  & 18   & 38  & 544 \\
                                & JSON Schema           & 432   & 11   & 269   & 18    & -     & -    & -   & -   \\
                                & RegEx                 & 448   & 9    & 0     & 272   & 1465  & 480  & 44  & 484 \\
                                & N/A                   & 361   & 24   & 312   & 40    & 1779  & 24   & 82  & 579 \\ \hline
\multirow{4}{*}{Qwen-2.5-7b-it} & TTMG                  & 476   & 64   & 25    & 165   & 1859  & 165  & 27  & 421 \\
                                & JSON Schema           & 433   & 41   & 242   & 13    & -     & -    & -   & -   \\
                                & RegEx                 & 444   & 15   & 0     & 270   & 1377     & 505    & 49   & 542   \\
                                & N/A                   & 408   & 4    & 304   & 23    & 1828  & 51   & 94  & 483 \\ \hline
\end{tabular}%
}
\vspace{-1em}
\end{table*}

\begin{tcolorbox}[leftrule=0mm,rightrule=0mm,toprule=0mm,bottomrule=0mm,left=0pt,right=0pt,top=0pt,bottom=0pt,title={RQ3: To what extent do existing structure-enforcing techniques, including strict template-based generation, reduce syntax errors, and what types of errors remain after syntax control?}]
    \textbf{Answer:} Template-based generation (TTMG) confirms that syntax errors can be fully controlled through strict enforcement. However, even under this idealized condition, substantial structural and semantic errors remain. This demonstrates that the core bottleneck of LLM structured output lies beyond syntax validity, pointing to unresolved challenges in structural adherence and semantic correctness.
\end{tcolorbox}

\section{Case Study}








To understand the practical implications of LLM structural and semantic failures, we conducted a qualitative analysis of the errors generated by the LLaMA-3.1-8B model on the BFCL v2 dataset. By examining the Abstract Syntax Tree (AST) matching logs, we can identify specific patterns in how the model breaks down when forced into structured formats.

\subsection{Qualitative Analysis for Structural Error}

\subsubsection{Error Classification}
The most frequent failure mode in structured output generation is the syntactic failure, predominantly flagged by the AST parser as \texttt{Invalid output format}. In our analyzed subset, this error alone accounted for 840 instances. These errors typically occur when the model wraps the output in incorrect delimiters or introduces unescaped characters, causing the parser to fail completely. For example, generating:
\begin{quote}
\texttt{[uber.ride(loc="2020 Addison Street, Berkeley, CA, USA", type="comfort", time=600)]}
\end{quote}
rather than adhering strictly to the expected JSON or standard Python object schema.

\subsubsection{Missing Branches}
Structural errors frequently manifest as incomplete generation, where the model successfully initiates a valid schema but fails to populate all required fields. We refer to this as ``Missing Branches.'' For instance, in media and weather domain tasks, the model frequently omitted mandatory arguments, producing errors such as:
\begin{quote}
\texttt{Missing required parameter 'genre'} (19 occurrences)\\
\texttt{Missing required parameter 'unit'} (15 occurrences)
\end{quote}
In these cases, the LLM successfully identifies the primary entity (e.g., the movie name or city) but neglects secondary parameters defined in the function signature.

\subsubsection{Wrong Key}
Another prevalent structural issue is the hallucination of keys or function names, where the model invents a structure not present in the provided schema (logged as \texttt{No ground truth for function}). For example, the model invoked \texttt{get\_weather} (11 times) or \texttt{Music\_3\_PlayMedia} instead of using the strictly defined API endpoints provided in the prompt. This indicates that the LLM occasionally ignores the provided context and defaults to its parameterized knowledge of common API names.

\subsection{Qualitative Analysis for Value Error}

\subsubsection{Error Classification}
Value errors occur when the model generates a perfectly valid structure (e.g., correct syntax) and includes all required keys, but the values assigned to those keys do not match the expected ground truth. These errors expose the semantic limitations and reasoning gaps of the model.

\subsubsection{Ambiguity}
Ambiguity errors happen when the model extracts a value that is colloquially correct but does not align with the strict canonical format expected by the downstream API. For example, when queried for a location, the model output:
\begin{quote}
\texttt{Parameter 'city' value 'SD, California'}
\end{quote}
whereas the strict ground truth was \texttt{['San Diego']}. Similarly, it generated \texttt{'Chi-town'} instead of \texttt{'Chicago, IL'}, and \texttt{'LA, CA'} instead of \texttt{'Los Angeles, CA'}. The model struggles with entity normalization without explicit formatting rules.

\subsubsection{Default Value}
LLMs frequently exhibit a ``temporal anchor'' bias, defaulting to their training cutoff or current operational year instead of the contextually provided temporal information. A stark example of this in our logs is date resolution errors: the model repeatedly generated values like:
\begin{quote}
\texttt{Parameter 'date' value '2024-03-12'}
\end{quote}
while the ground truth expectation based on the prompt's context was \texttt{'2023-03-12'}. The model defaults to the year 2024 rather than anchoring to the temporal constraints embedded in the task.

\subsubsection{Miss Understanding}
Misunderstanding errors occur when the model fails to comprehend the scale, type, or specific domain terminology of a parameter. A prime example from the dataset is a financial parameter misinterpretation: the model output \texttt{Parameter 'amount' value '155.0'}, but the API expected the value in cents, \texttt{[15500.0]}. Another example is domain categorization, where the model assigned \texttt{Parameter 'genre' value 'Hillbilly'}, missing the broader target classification \texttt{['Country']}. 

\subsubsection{Failure Output}
Failure output represents cases where the model populates a value field with conversational text or truncated strings instead of the precise target entity. For example, instead of a concise string, the model output a full conversational paragraph:
\begin{quote}
\texttt{Parameter 'Content' value 'The recent breakthroughs in machine learning algorithms have led to significant advancements...'}
\end{quote}
failing to extract just the specific target sentence requested by the system prompt.

\subsection{Suggestion for Model Side}

Based on the qualitative analysis of these errors, several improvements can be made on the model side for structure-aware SE tasks:
\begin{enumerate}
    \item \textbf{Schema-Constrained Fine-Tuning:} To combat ``Missing Branches'' and ``Invalid output format'' errors, models should be fine-tuned with a heavier emphasis on strict schema adherence. Techniques like constrained decoding should ideally be integrated natively into the fine-tuning loss to penalize the omission of required keys.
    \item \textbf{Context Anchoring:} To address ``Ambiguity'' and ``Default Value'' errors, models need explicit training to prioritize context-provided temporal markers (like base dates) and entity normalization formats over their pre-trained world knowledge.
    \item \textbf{Type and Scale Awareness:} For APIs expecting specific data types or scaled values (e.g., cents vs. dollars), models require improved arithmetic reasoning and type-awareness during value generation to prevent critical downstream execution failures.
\end{enumerate}

\section{Discussion}

Our study highlights that structural fidelity is a pivotal challenge for deploying LLMs in software engineering tasks. 
Through large-scale evaluation, we observed that structural and syntactic errors frequently prevent outputs from being directly usable. 
While grammar- and regex-based techniques reduce many format violations, they do not guarantee complete validity. 
By employing TTMG as a strict control, we confirmed that syntax errors can be virtually eliminated. 
However, this did not lead to perfect outputs: substantial structural and logical errors remained, such as missing required fields, schema mismatches, or semantically incorrect values.

This finding underscores a key insight: \textbf{syntax is not the root bottleneck}. 
Even when models are forced into perfectly valid templates, they often fail to populate placeholders with the correct content. 
Thus, many of the reliability issues in structured outputs stem from deeper structural or semantic misalignments rather than surface-level formatting mistakes. 
Our results align with prior studies: enforcing a target structure (e.g., via JSON schemas or function calling) is effective at eliminating format violations, but offers only limited gains in correctness~\cite{geng2025json}. 
This reflects a broader trend—format control alone does not guarantee better reasoning or factual accuracy. 
Moreover, overly strict constraints may even impair the model’s reasoning ability in complex tasks, as noted by Tam et al.~\cite{tam2024let}.

For practitioners, these results carry two implications. 
First, syntax validation tools (e.g., schema checking, regex constraints, or template filling) are necessary to prevent trivial parsing errors and ensure outputs can be consumed by downstream systems. 
Second, however, these tools are insufficient for achieving reliable end-to-end correctness. 
Semantic validation and reasoning-aware mechanisms remain necessary, especially in high-stakes workflows such as code generation, API construction, or financial systems.

For researchers, the controlled TTMG experiments provide an important diagnostic: 
they separate \textit{what errors can be solved by stricter syntax control} from \textit{what errors persist regardless of syntactic enforcement}. 
This suggests that future work should move beyond grammar- or template-level enforcement and instead focus on methods that integrate structural fidelity with semantic understanding and reasoning.

\paragraph{Structural Fidelity vs.\ Benchmark Memorization.}
These findings also invite a comparison with a superficially related concern that has received considerable attention in the LLM literature: benchmark memorization. Memorization is primarily an evaluation-integrity problem—when a model has seen benchmark instances during training, its reported scores become inflated and no longer reflect genuine capability, making cross-model comparisons unreliable \cite{cheng2025survey,dekoninck2024constat,palavalli2024taxonomy}. Crucially, however, memorization is largely a \emph{measurement} artefact: in a real deployment, if a model produces functionally correct code for a given task—whether it ``remembers'' the solution or derives it afresh—the practitioner's objective is satisfied. The code works, and that is what matters in practice.

Structural fidelity, by contrast, is an \emph{operational} concern that directly determines whether an output is usable at all. Even a semantically perfect solution is worthless if it cannot be consumed by the surrounding infrastructure. Our own results illustrate this sharply: under TTMG, which eliminates virtually all syntax errors, LLaMA-3.1-8B still produces 56 structural and 185 value errors on CallNavi, and 38 structural and 544 value errors on BFCL~v2 (Table~\ref{tab:templated}). The model demonstrably ``knows'' the correct action—AST match rates improve from 361 to 462 on CallNavi once syntax noise is removed—yet hundreds of outputs remain unusable because they violate the structural contract expected by the consuming system. Similarly, the unconstrained results in Table~\ref{tab:nonstruct-result} show that even GPT-4.1-mini, which achieves the highest scores on most tasks, still produces 297 structural errors on CallNavi and 636 on BFCL~v2, confirming that scale and capability do not by themselves resolve the structural fidelity gap.

In modern software engineering practice, isolated code generation is the exception rather than the norm: outputs must slot into build systems, API gateways, data pipelines, and CI/CD workflows that impose rigid format contracts \cite{xu2025logsage}. A model that consistently solves the reasoning problem but intermittently violates these structural contracts introduces a class of failure that is invisible to conventional accuracy metrics yet critical in production. Unlike memorization, which inflates perceived ability without affecting real-world utility, structural non-compliance renders genuinely correct solutions unusable. Structural fidelity is therefore not merely an academic critique of model behaviour but a prerequisite for practical deployment—one that persists regardless of whether the underlying solution was memorized or genuinely reasoned \cite{margiotta2025evaluating}.

\paragraph{Implications of Mixture-of-Experts Architectures.}
The results from our two MoE models: Qwen3-30B-A3B and Mixtral-8x7B-IT (Tables~\ref{tab:nonstruct-result} and~\ref{tab:structresult}) reveal a consistent pattern that warrants careful attention. Despite achieving task-level performance on par with GPT-4.1-mini, Qwen3-30B-A3B proved largely impervious to the structure-enforcing tools that substantially help dense models: BFCL~v2 syntax errors dropped only from 134 to 123 with Outlines (versus 1,833 to 2 for LLaMA-3.1-8B under the same tool), and structural errors remained static at 622. Mixtral-8x7B-IT reinforces this finding even more sharply: under LLGuidance, its BFCL~v2 AST match \emph{collapsed} from 0.4086 (unconstrained) to 0.2363, meaning the tool actively degraded performance rather than improving it while structural errors remained high at 408. On CallNavi, both tools left Mixtral's exact match effectively unchanged at $\sim$0.41 with $\sim$340 structural errors. We attribute this to the fundamental mismatch between where current tools intervene, i.e. at the output logit level, and where MoE architectures make their critical decisions, i.e., at the expert-routing layer in the intermediate blocks. Because the sparse gating mechanism has already committed to a particular subset of experts before logit-level constraints are applied, post-hoc token masking cannot redirect the generation trajectory that produced the structural violation.

To appreciate the scale of this challenge, consider the internal structure of these models: Qwen3-30B-A3B activates only 8 of 128 experts per token (3.3B of 30.5B total parameters), while Mixtral-8x7B-IT activates 2 of 8 experts per token. In both cases, the structural ``character'' of each generated token is determined by \emph{which} experts the router selects, a decision entirely opaque to current logit-level tools. The dominant trend in frontier LLM development is toward precisely this kind of sparse architecture: DeepSeek-V3 and the broader Qwen3 MoE family all follow the same paradigm. Our results across both MoE models, where Qwen3 shows tool-resistance, and Mixtral shows active degradation, suggest that this is not an isolated quirk but a systematic limitation. If structure-enforcing tools are fundamentally limited in their ability to steer sparse architectures, the structural fidelity problem is poised to grow more acute as the field advances.

These findings point toward a concrete research direction: structure-aware interventions that operate \emph{within} the model, not merely at its output. One promising avenue is to incorporate a dedicated \emph{structure-following expert}: a specialist sub-network within the MoE layer explicitly trained on format-adherence objectives and to bias the router toward activating it whenever the generation context requires structural compliance. More broadly, neurosymbolic approaches that embed grammar or schema constraints directly into the routing or attention mechanism could provide the architectural hook that logit-level masking lacks. Exploring these directions is, in our view, essential for achieving reliable structural control over the next generation of MoE-based LLMs.

\paragraph{Implications of Mixture-of-Experts Architectures.}
Our evaluation of Qwen3-30B-A3B (Tables~\ref{tab:nonstruct-result} and~\ref{tab:structresult}) reveals a pattern that warrants careful attention. Despite achieving task-level performance on par with GPT-4.1-mini, this model proved largely impervious to the structure-enforcing tools that substantially help dense models. We attribute this to the fundamental mismatch between where current tools intervene: at the output logit level, and where the MoE architecture makes its critical decisions: at the expert-routing layer in the intermediate blocks. Because the sparse gating mechanism has already committed to a particular subset of experts before logit-level constraints are applied, post-hoc token masking cannot redirect the generation trajectory that produced the structural violation in the first place.

To appreciate the scale of this challenge, consider the internal structure of Qwen3-30B-A3B: each of its 48 transformer layers contains 128 experts, of which only 8 are activated per token via a learned softmax router \cite{qwen3technicalreport}. The model thus holds 30.5B total parameters but activates only 3.3B per token, a ratio that maximises capacity while minimising inference cost. However, it also means that the structural ``character'' of each generated token is determined by \emph{which} 8-out-of-128 experts the router selects, a decision that is entirely opaque to current logit-level tools. The dominant trend in frontier LLM development is toward precisely this kind of sparse architecture: Mixtral, DeepSeek-V3, and the broader Qwen3 MoE family \cite{jiang2024mixtral,liu2024deepseek,qwen3technicalreport} all follow the same paradigm. If structure-enforcing tools are fundamentally limited in their ability to steer sparse architectures, the structural fidelity problem is poised to grow more acute, not less, as the field advances.

Our study evaluates only a single MoE model and thus cannot claim generality, but the consistent ineffectiveness of \emph{all} tested tools on this architecture points toward a concrete research direction: structure-aware interventions that operate \emph{within} the model, not merely at its output. One promising avenue is to incorporate a dedicated \emph{structure-following expert}, a specialist sub-network within the MoE layer that is explicitly trained or fine-tuned on format-adherence objectives, and to bias the router to activate it whenever the generation context requires structural compliance. More broadly, neurosymbolic approaches that embed formal grammar or schema constraints directly into the routing or attention mechanism could provide the architectural hook that logit-level masking lacks. Such methods would allow the model to ``plan'' structurally at the same stage where it plans semantically, rather than relying on an external post-hoc correction that arrives too late in the computation to redirect expert selection. Exploring these directions is, in our view, essential for achieving reliable structural control over the next generation of MoE-based LLMs.

In summary, our results show that while syntax enforcement is an important first step, the core challenge of reliable LLM structured output lies in resolving deeper structural and semantic errors. 
This distinction clarifies the limitations of current approaches and highlights promising directions for more robust solutions.

\section{Related Work}

Since the LLM structure output is an underexplored, experimental topic in the AI-SE domain, we do not have much previous work to review. Controlling structured outputs from large language models (LLMs) has become critical in software engineering tasks, where syntactic accuracy is essential for downstream processing \cite{geng2025json}. Grammar-constrained decoding methods, such as PICARD \cite{scholak-etal-2021-picard} and Grammar-Constrained Decoding (GCD) \cite{geng2023grammar}, directly enforce syntactic validity during token generation, significantly reducing formatting errors in tasks like SQL generation. Similarly, grammar prompting approaches leverage minimal grammars embedded in prompts to guide structured generation, offering competitive performance without model fine-tuning \cite{wang2023grammar}.

JSON schema and regex-based constraint methods have also gained attention, particularly through general frameworks such as Guidance \cite{guidance2023}, Outlines \cite{willard2023efficient}, and XGrammar \cite{dong2024xgrammar}. These methods translate user-defined schemas into token-level generation constraints, effectively reducing syntax errors with modest computational overhead. Additionally, ReLM \cite{kuchnik2023validating} employs regex-based validation during decoding, enabling token-level backtracking to ensure strict format adherence. The introduction of features like OpenAI's function calling with structured JSON responses further demonstrates practical schema-based output constraints \cite{tam2024let}. However, recent studies note that overly strict format restrictions can negatively impact the semantic quality of LLM outputs.

\section{Threats To Validity}
Like any empirical study, our work is subject to several threats to validity.

\textbf{Internal validity.} 
Our experiments rely on specific implementations of grammar-constrained decoding, regex validation, and template-based generation (TTMG). 
Although TTMG was introduced here, it is not positioned as a novel contribution but as a strict control condition. 
This means that the design of our template-filling processor directly impacts the results: if the implementation contained bugs or overly permissive transitions, the conclusions about residual errors might be affected. 
To mitigate this, we carefully verified correctness using deterministic tests on synthetic templates and confirmed that TTMG always produces syntactically valid outputs. 
Still, our classification of errors (syntax, structural, semantic) involves some judgment and may not capture all nuances.

\textbf{External validity.} 
We evaluated four representative software engineering tasks, but the findings may not generalize to all forms of structured output. 
For example, tasks involving deeply nested schemas, complex control flow, or multimodal contexts may exhibit different error distributions. 
Moreover, we used two open-source LLMs (LLaMA-3.1-8B and Qwen-2.5-7B-it); results could differ with larger proprietary models or with domain-specific fine-tuning. 
Nevertheless, the key insight—that syntax enforcement can be controlled, yet deeper structural and semantic errors persist—is likely to hold across architectures, since it reflects fundamental limitations of current LLM reasoning and alignment.

\textbf{Construct validity.} 
Our metrics focus on syntax errors, structural violations, and semantic mismatches measured via AST matching and error counts. 
While these metrics capture important aspects of structured fidelity, they may not fully reflect user-facing correctness. 
For example, an output could be structurally valid but still useless for a downstream task due to domain-specific mistakes. 
We partially addressed this by conducting a case study that illustrates practical impacts of structural errors, but additional human evaluation could strengthen the validity of our conclusions.

\textbf{Limitations of TTMG as a control.} 
While TTMG provides a valuable diagnostic setting by eliminating syntax errors, it is not universally applicable. Template-driven decoding requires a fully specified structure in advance, which may not exist for tasks with dynamic or partially known schemas. In scenarios with highly flexible outputs, such as natural language explanations or open-domain dialogues, strict template filling is infeasible.  Furthermore, even in structured tasks, complex nested templates may become cumbersome to construct and maintain. Thus, TTMG should be viewed as a controlled experimental tool rather than a general-purpose solution, helping to isolate error sources but not necessarily reflecting realistic deployment conditions.

In summary, our study’s limitations mainly arise from the chosen tasks, models, and evaluation metrics.  However, because TTMG was used as a strict control condition, we believe the central conclusion—that syntax errors can be eliminated, but structural and semantic challenges remain—stands robust to these validity concerns.

\section{Conclusion}
\subsection{Practical Conclusion}
For software engineering practitioners deploying LLMs, the primary takeaway is that \textbf{syntax is no longer the ultimate bottleneck}. By utilizing strict structure-enforcing tools---such as template-driven generation (TTMG), regex validators, or grammar-constrained decoding---practitioners can effectively drive superficial parsing failures (like malformed JSON) to near zero. 

However, current grammar-constrained and rule-based generation tools still have a long way to go to be fully practical as standalone solutions. As our empirical evaluation and case study demonstrate, forcing an LLM into a syntactically perfect mold does not cure its underlying reasoning deficits. Eliminating syntax errors merely exposes deeper, unresolved challenges: structural hallucinations (such as omitting required keys) and semantic inaccuracies (such as populating fields with incorrect or default values). 

Therefore, in real-world SE deployments, practitioners should treat structure-enforcing tools as a mandatory baseline defense to prevent pipeline crashes, but must actively pair them with robust semantic validation, type-checking, and context-anchored prompting to achieve true end-to-end reliability.

\subsection{Overall}
In this work, we systematically evaluated the challenges of structured output generation in LLMs across several software engineering tasks. Our large-scale analysis showed that structural and syntactic errors are a major obstacle, frequently preventing LLM outputs from being directly usable. By benchmarking grammar-constrained decoding, regex-based validation, and template-based generation (TTMG), we found that syntax-level errors can largely be eliminated under strict enforcement. However, our results demonstrate that deeper issues—such as missing fields, schema mismatches, and logical inconsistencies—remain unsolved even when syntax is fully controlled.

These findings indicate that the core bottleneck is not simply syntax compliance but the model’s difficulty in consistently aligning structure with semantics. While current tools provide value by reducing format violations, their improvements in overall task accuracy remain limited. For practitioners, this suggests that deploying LLMs in high-stakes workflows requires not only syntax validation but also complementary mechanisms to ensure semantic correctness. For software engineering researchers, our study provides an empirical basis to move beyond surface-level format enforcement and toward approaches that integrate structural fidelity with deeper reasoning reliability.

\section*{Declarations}
\paragraph*{Funding: } The FNR(Luxembourg National Research Fund) funded this research under grants NCER22/IS/16570468/NCERFT.
\paragraph*{Ethical approval: } not applicable.
\paragraph*{Informed consent: } not applicable.
\paragraph*{Author Contributions: } Conceptualization: YS, JK; Methodology: YS, SE; Software: YS; Validation: YS, PR; Formal analysis: YS, PR; Investigation: YS; Writing - Original Draft: YS, PR; Writing - Review \& Editing: YS, PR, TS, JK; Visualization: YS; Supervision: JK, TB; Project administration:
JK; Funding acquisition: JK, TB.
\paragraph*{Data Availability Statement: }
This article is licensed under a Creative Commons Attribution 4.0 International License, which permits use, sharing, adaptation, distribution and reproduction in any medium or format, as long as you give appropriate credit to the original author(s) and the source, provide a link to the Creative Commons licence, and indicate if changes were made. The images or other third party material in this article are included in the article’s Creative Commons licence, unless indicated otherwise in a credit line to the material. If material is not included in the article’s Creative Commons licence and your intended use is not permitted by statutory regulation or exceeds the permitted use, you will need to obtain permission directly from the copyright holder. To view a copy of this licence, visit http://creativecommons.org/licenses/by/4.0/.
\paragraph*{Conflict of Interest: }This project is under technology collaborate with BGL BNP PARIBAS. The authors have no other relevant financial or non-financial interests to disclose.

\section*{Appendix}

\section*{Algorithms for TTMG}
\label{apd:alg}
Formally, let the template be represented as an alternating sequence of literal token sequences $s_0, s_1, \ldots, s_N$ and placeholders $\square_1, \square_2, \ldots, \square_N$, for example $T = s_0,\square_1,s_1,\square_2,\cdots,\square_N,s_N$. This template can be seen as defining a regular language (a very strict subset of a context-free grammar) that the output must match. We construct an FSM $M = (Q, q_0, q_F, \Sigma, \delta)$ for $T$, where each state $q_i \in Q$ corresponds to a position in the template (e.g. “currently filling placeholder $\square_i$” or “about to emit literal $s_i$”). The transition function $\delta$ encodes the allowed next token(s) given the current state. If the next segment is a fixed literal $s_i = [w_1, w_2, \ldots, w_k]$, the FSM will deterministically require those tokens in order: $\delta(q, w_1) = q',, \delta(q', w_2) = q'', \dots$ until the literal is fully matched, transitioning through a sequence of intermediate states for each forced token. Conversely, if the current state corresponds to a placeholder $\square_i$, $\delta$ permits a transition on any token that is valid within that free-form slot (typically almost any vocabulary token, possibly excluding ones that would prematurely terminate the slot). We denote by $\mathcal{A}(q)$ the set of tokens allowed in state $q$: for a literal state expecting $w$, $\mathcal{A}(q) = {w}$; for a placeholder state, $\mathcal{A}(q)$ would be the full vocabulary (or a large subset thereof) since any token is in principle allowed as content. At each decoding step $t$, the processor modifies the language model’s logits $z_t$ by masking out any token not in $\mathcal{A}(q)$ (setting its logit to $-\infty$). This yields a restricted distribution $P(y_t \mid y_{<t},,q)$ concentrated only on valid next tokens $y_t \in \mathcal{A}(q)$. Sampling or greedy selection is then applied to this filtered distribution to obtain the next token $y_t$. Finally, the FSM state is updated $q \leftarrow \delta(q, y_t)$, advancing to the next state. This process repeats until the FSM reaches the final state $q_F$, indicating the entire template (all literal segments and placeholders) has been generated. The pseudocode of this procedure is shown in the Algorithm 
\ref{alg:ttmg}.

\algtext*{EndIf}
\begin{small}

\begin{algorithm}
\caption{Template Token Match Generation (TTMG)}
\label{alg:ttmg}
\small
\begin{algorithmic}[1]
\State Initialize $c \leftarrow 1$, $p \leftarrow 0$, $\text{mode} \leftarrow \text{FORCED}$
\Procedure{GenerationStep}{input\_ids, scores}
    \If{$\text{length}(\text{input\_ids}) \leq \mathcal{P}$}  \Comment{$\mathcal{P}$: prompt prefix length}
        \State \Return scores \Comment{Do not modify prompt}
    \EndIf

    \If{$\text{mode} == \text{FORCED}$}
        \State segment $\leftarrow S[c]$
        \If{$p < |\text{segment}|$}
            \State last\_token $\leftarrow \text{input\_ids}[-1]$
            \State expected\_token $\leftarrow \text{segment}[p]$
            \If{last\_token $==$ expected\_token}
                \State $p \leftarrow p + 1$
                \If{$p == |\text{segment}|$}
                    \State mode $\leftarrow \text{FREE}$
                \EndIf
            \EndIf
        \EndIf
        \If{$\text{mode} == \text{FORCED}$ \textbf{and} $p < |\text{segment}|$}
            \State scores[:] $\leftarrow -\infty$
            \State scores[expected\_token] $\leftarrow$ \texttt{FORCE\_SCORE}
            \State \Return scores
        \EndIf
    \EndIf

    \If{$\text{mode} == \text{FREE}$}
        \If{$c < n$}
            \State next\_segment $\leftarrow S[c+1]$
            \State last\_token $\leftarrow \text{input\_ids}[-1]$
            \If{last\_token $==$ next\_segment[0]}
                \State $c \leftarrow c + 1$
                \State $p \leftarrow 1$ \Comment{First token of next segment already matched}
                \State mode $\leftarrow \text{FORCED}$
                \If{$p < |\text{next\_segment}|$}
                    \State expected\_token $\leftarrow \text{next\_segment}[p]$
                    \State scores[:] $\leftarrow -\infty$
                    \State scores[expected\_token]$\leftarrow$ \texttt{FORCE\_SCORE}
                    \State \Return scores
                \EndIf
            \EndIf
        \EndIf
    \EndIf
    \State \Return scores \Comment{Unconstrained generation step}
\EndProcedure
\end{algorithmic}
\end{algorithm}
\end{small}

This procedure alternates between forcing deterministic tokens and allowing free-form generation in a single decoding pass. In terms of theoretical complexity, this procedure has a negligible performance impact. Intuitively, when the decoder is in a state expecting a literal segment $s_i$, exactly the next token of $s_i$ will be available and chosen (all others being masked out). This effectively injects the template’s required text into the output at the proper positions without sampling. Conversely, when in a placeholder state, the processor lifts constraints and permits the model to freely generate tokens until the placeholder is complete. A placeholder may span an arbitrary number of tokens (up to some predefined limit or until it naturally transitions when the next literal’s first token is produced). In practice, the FSM can be designed so that encountering the next literal’s initial token triggers a state transition out of the placeholder, thereby cleanly handing control back to a forced segment. This guarantees that as soon as the model has produced content for the slot, the decoding will snap back to the rigid template for the next portion. By construction, any output $Y$ emitted by this process will exactly match the template pattern, with no extraneous tokens and no missing fixed tokens. In other words, the TemplateFillingProcessor ensures strong structural compliance: it’s not just syntactically well-formed, but conforms to an exact format (down to specific words and punctuation in $T$) that may be required by the task.

\bibliographystyle{spmpsci}      
\bibliography{custom}   

\end{document}